\documentclass[prc,preprint,showpacs,showkeys,
tightenlines,
        amsfonts,nofootinbib]{revtex4}

\usepackage{graphicx}  %
\usepackage{bm}  %
\usepackage{amssymb}
\usepackage{dcolumn}

\begin{document}

%


\newcommand{\beq}{\begin{equation}}
\newcommand{\eeq}{\end{equation}}
\newcommand{\bea}{\begin{eqnarray}}
\newcommand{\eea}{\end{eqnarray}}
\newcommand{\ben}{\begin{eqnarray*}}
\newcommand{\een}{\end{eqnarray*}}

\newcommand{\simlt}{\stackrel{<}{{}_\sim}}
\newcommand{\simgt}{\stackrel{>}{{}_\sim}}
\newcommand{\sing}{$^1\!S_0$ }
\newcommand{\btau}{\mbox{\boldmath$\tau$}}
\newcommand{\bsig}{\mbox{\boldmath$\sigma$}}

\newcommand{\dt}{\partial_t}

\newcommand{\kf}{k_{\rm F}}
\newcommand{\wt}{\widetilde}
\newcommand{\kt}{\widetilde k}
\newcommand{\pt}{\widetilde p}
\newcommand{\qt}{\widetilde q}
\newcommand{\wh}{\widehat}
\newcommand{\dens}{\rho}
\newcommand{\edens}{{\cal E}}
\newcommand{\order}[1]{{\cal O}(#1)}

\newcommand{\psihat}{\widehat\psi}
\newcommand{\xvec}{{\bf x}}
\newcommand{\dagphan}{{\phantom{\dagger}}}
\newcommand{\kvec}{{\bf k}}
\newcommand{\kpvec}{{\bf k}'}
\newcommand{\ak}{a^\dagphan_\kvec}
\newcommand{\akdag}{a^\dagger_\kvec}
\newcommand{\akv}[1]{a^\dagphan_{\kvec_{#1}}}
\newcommand{\akdagv}[1]{a^\dagger_{\kvec_{#1}}}
\newcommand{\akp}{a^\dagphan_{\kvec'}}
\newcommand{\akpdag}{a^\dagger_{\kvec'}}
\newcommand{\akpv}[1]{a^\dagphan_{\kvec'_{#1}}}
\newcommand{\akpdagv}[1]{a^\dagger_{\kvec'_{#1}}}

\def\vec#1{{\bf #1}}

\newcommand{\nab}{\overrightarrow{\nabla}}
\newcommand{\nabsq}{\overrightarrow{\nabla}^{2}\!}
\newcommand{\nabl}{\overleftarrow{\nabla}}
\newcommand{\galnab}{\tensor{\nabla}}
\newcommand{\psid}{{\psi^\dagger}}
\newcommand{\psidal}{{\psi^\dagger_\alpha}}
\newcommand{\psidbe}{{\psi^\dagger_\beta}}
\newcommand{\idt}{{i\partial_t}}
\newcommand{\Sthree}{{\delta_{11'}(\delta_{22'}\delta_{33'}%
        -\delta_{23'}\delta_{32'})%
        +\delta_{12'}(\delta_{23'}\delta_{31'}-\delta_{21'}\delta_{33'})%
        +\delta_{13'}(\delta_{21'}\delta_{32'}-\delta_{22'}\delta_{31'})}}
\newcommand{\Stwo}{{\delta_{11'}\delta_{22'}-\delta_{12'}\delta_{21'}}}
\newcommand{\Left}{{\cal L}}
\newcommand{\Tr}{{\rm Tr}}

\newcommand{\h}{\hfil}
\newcommand{\be}{\begin{enumerate}}
\newcommand{\ee}{\end{enumerate}}
\newcommand{\I}{\item}   

\newcommand{\density}{\rho}

\newcommand{\thyp}{\mbox{---}}

\newcommand{\Jks}{J_{\ks}}
\newcommand{\Jzero}{\Jks}
\newcommand{\Jdensityzero}{J_\density^0}

\newcommand{\ks}{{\rm ks}}
\newcommand{\Seq}{Schr\"odinger\ equation }
\newcommand{\yvec}{{\bf y}}
\newcommand{\ve}{V_{eff}}
\newcommand{\densityJzero}{\density^0_J}
\newcommand{\Dfunct}{{D^{-1}}}
\newcommand{\drv}[2]{{\mbox{$\partial$} #1\over \mbox{$\partial$} #2}}
\newcommand{\drvs}[2]{{\mbox{$\partial^2$} #1\over \mbox{$\partial$} #2 \mbox{$^2$}}}
\newcommand{\drvt}[2]{{\partial^3 #1\over \partial #2 ^3}}
\newcommand{\til}[1]{{\widetilde #1}}
\newcommand{\dthreex}{d^3\xvec}
\newcommand{\dthreey}{d^3\yvec}

\newcommand{\efermi}{\varepsilon_{{\scriptscriptstyle \rm F}}}
\newcommand{\eHF}{\wt\varepsilon}
\newcommand{\eKS}{e}
\newcommand{\ekJ}{e_\kvec^J}
\newcommand{\epsk}{\varepsilon_\kvec}
\newcommand{\epsKS}{\varepsilon}
\newcommand{\Eq}[1]{Eq.~(\ref{#1})}

\newcommand{\Fi}[1]{\mbox{$F_{#1}$}}
\newcommand{\fq}{f_{\qvec}}

\newcommand{\Gammaalt}{\overline\Gamma}
\newcommand{\Gammaks}{\Gamma_{\ks}}
\newcommand{\Gamint}{\widetilde{\Gamma}_{\rm int}}
\newcommand{\GKS}{G_{\ks}}
\newcommand{\grad}{{\bm{\nabla}}}   
\newcommand{\greenKS}{{G}_{\ks}}
\newcommand{\intint}{\int\!\!\int}

\newcommand{\kfermi}{k_{{\scriptscriptstyle \rm F}}}   
\newcommand{\kJzero}{k_J}

\renewcommand{\l}{\lambda}

\newcommand{\MeV}{\mbox{\,MeV}}
\newcommand{\mi}[1]{\mbox{$\mu_{#1}$}}

\newcommand{\Oi}[1]{\mbox{$\Omega_{#1}$}}

\newcommand{\phibar}{\overline\phi}
\newcommand{\phidagger}{\phi^\dagger}
\newcommand{\phistar}{\phi^\ast}
\newcommand{\psibar}{\overline\psi}
\newcommand{\psidagger}{\psi^\dagger}
\newcommand{\qvec}{\vector{\rho}}

\newcommand{\tr}{{\rm tr\,}}

\newcommand{\Ulong}{U_{L}}

\renewcommand{\vector}[1]{{\bf #1}}
\newcommand{\vext}{v_{\rm ext}}   
\newcommand{\Vlong}{V_{L}}

\newcommand{\Wzero}{W_0}
\newcommand{\Wks}{W_{\ks}}
\newcommand{\zvec}{\vector{z}}
\newcommand{\rvec}{\vector{r}}

%
%
%
%


\title{Density Functional Theory for
   a Confined Fermi System with Short-Range Interaction}

\author{S.J.\ Puglia}\email{puglia@campbell.mps.ohio-state.edu}
\author{A. Bhattacharyya}\email{anirban@mps.ohio-state.edu}
\author{R.J. Furnstahl}\email{furnstahl.1@osu.edu}
\affiliation{Department of Physics,
         The Ohio State University, Columbus, OH\ 43210}

%
\date{December, 2002}

\begin{abstract}
Effective field theory (EFT) methods are applied to density functional
theory (DFT) as part of a program
to systematically go beyond mean-field approaches to medium and heavy nuclei.
A system of fermions with short-range, natural interactions and an external
confining potential (e.g., fermionic atoms in an optical trap) serves as a
laboratory for studying DFT/EFT.
An effective action formalism leads to a Kohn-Sham DFT 
by applying an inversion method
order-by-order in the EFT expansion parameter.
Representative results showing the convergence of Kohn-Sham calculations
at zero temperature 
in the local density approximation (LDA) are compared to Thomas-Fermi
calculations and to power-counting
estimates.
\end{abstract}

\smallskip
\maketitle


\section{Introduction}
\label{sec:introduction}

  Calculations of bulk observables for medium to heavy nuclei
typically rely on nonrelativistic (Skyrme) or covariant (QHD)
mean-field models. How can one systematically go beyond such models? 
One possibility is to view nuclear mean-field approaches as approximate
implementations of 
Kohn-Sham density functional theory 
(DFT) \cite{KOHN65,PARR89,DREIZLER90,KOHN99,ARGAMAN00}, 
which is widely used in condensed matter and quantum chemistry applications.
To date,
the refinement of DFT methods has focused
almost exclusively on Coulomb systems
and DFT has had little explicit impact on
nuclear structure phenomenology \cite{PETKOV91}.
Effective field theory
(EFT), however, offers a systematic approach for describing 
low-energy nuclear physics that can provide a framework for nuclear DFT. 
EFT approaches have been making steady progress on two- and 
three-body nuclear systems and certain halo nuclei, and {\it ab initio\/}
shell model methods should be able to extend the calculations to a wide
range of light 
nuclei \cite{EFT98,EFT99,Birareview,BEANE99,Furnstahl:2001hs}. 
Our ultimate goal is to systematically describe
\emph{heavier} nuclei by applying EFT methods to Kohn-Sham
DFT.  In this paper we take the first steps toward this goal.

Density functional theory  provides a calculational framework that
greatly extends the range of many-body calculations in finite systems.
In Kohn-Sham DFT, the coordinate-space
ground-state density for $A$ fermions is found by simply summing the
squared wavefunctions for a set of single-particle orbitals. These
wavefunctions are solutions to a Schr\"odinger equation featuring the 
Kohn-Sham single-particle potential, which is local and energy
independent.  
At zero temperature, for nondegenerate ground states (e.g., closed
shells),
the sum for the density is over the orbitals with the
lowest $A$ eigenvalues, equally
weighted (i.e., ``occupation numbers'' are either one or zero).
The Kohn-Sham potential is itself a functional of the
orbitals, so we have a self-consistent problem that is solvable by
iteration.   When the iterations have converged, the orbitals can be
plugged into an energy functional to find the ground-state energy.  
While this procedure is very similar to the self-consistent
Hartree approximation, 
it incorporates \emph{all} correlations \emph{if} the correct Kohn-Sham 
functional is used.
 
The advantages of the Kohn-Sham DFT procedure are clear.  
One is that the external potential, if present, appears in a very simple
way in a separate term in the functional, with the rest 
of the functional being universal.
Another is that solving 
the Schr\"odinger equation with local potentials for eigenvalues is 
relatively simple and fast.  
The computational advantages would be lost, of course, if the construction and
evaluation of the Kohn-Sham potential is itself too difficult
or expensive. 
The experience for
Coulomb systems is that local density approximations (LDA) for the potential
part of the energy functional  and
their improvement with gradient expansions
work very well in many (although not all) systems
\cite{PERDEW96,PERDEW99,ARGAMAN00}. 
That is, the most important nonlocality to treat explicitly is the
kinetic energy, which is why the Kohn-Sham framework is generally superior to
the Thomas-Fermi approximation,
in which the kinetic energy is given by an LDA
(see Sect.~\ref{subsec:TF} below).
The gradient expansion approximations greatly simplify the
evaluation of the Kohn-Sham potential. 
We anticipate that the convergence of
derivative expansions for nuclear systems will be rapid in general,
but this will need to be confirmed. 

Our strategy is to 
apply an effective action formalism \cite{COLEMAN88,PESKIN95,WEINBERG96}
to calculate the Kohn-Sham
potential and energy functional order-by-order in an EFT expansion,
and to use EFT power counting to organize and justify a derivative
expansion of the functional.  
In the present work, we use a dilute, confined Fermi system with
short-range interactions as a 
laboratory to explore  how EFT can be used to carry out
systematic DFT calculations.
In this case, the explicit
expansion parameter is the local Fermi momentum times the
scattering length (and other effective range parameters).
 We assume 
a gradient expansion parameter that justifies a local
density approximation, but
the verification of this assumption is postponed to future
work.
Ultimately we are interested in calculating self-bound systems (e.g.,
nuclei), with spin- and isospin-dependent interactions and long-range
forces (e.g., pion exchange).
These are all significant but well-defined
extensions of the model described here.
In the meantime, the model provides a prototype for more complex systems
and also has a physical realization in recent and forthcoming experiments on
fermionic atoms in optical traps \cite{GRANADE02}.

The Kohn-Sham approach to DFT was proposed in Ref.~\cite{KOHN65}.
Since then, the literature of DFT
applications has grown exponentially, 
primarily in the areas of quantum chemistry and
electronic structure \cite{ARGAMAN00}.  
A general introduction to density functional theory as conventionally
applied is provided in the books by Dreizler and Gross \cite{DREIZLER90}
and Parr and Young \cite{PARR89}, 
while Ref.~\cite{KOCH00} is a  practitioners guide to DFT for quantum
chemists.
The connection of DFT to nonrelativistic mean-field approaches 
to nuclei (e.g.,
Skyrme models) was pointed out in Ref.~\cite{BRACK85} (and no doubt
elsewhere) and
was explored for covariant nuclear mean-field models in 
Refs.~\cite{SCHMID95,SCHMID95a}.
However, it has not led, to our knowledge, to new or systematically
improved mean-field-type functionals for nuclei.

The use of functional Legendre transformations for DFT with the effective
action formalism was first
detailed by Fukuda and collaborators
\cite{FUKUDA94,FUKUDA95}, who also discuss the inversion and auxiliary
field methods of constructing the effective action.
The connection to Kohn-Sham DFT was shown by Valiev and Fernando
\cite{VALIEV96,VALIEV97,VALIEV97b,RASAMNY98} and later by other authors in
Refs.~\cite{FAUSSURIER00,CHITRA00,CHITRA01}. 
Recent work by Polonyi and Sailer applies renormalization group methods 
and a cluster expansion to
an effective-action formulation of generalized 
DFT for Coulomb systems \cite{Polonyi:2001uc}.
To our knowledge, however,
there is no prior work on merging the Kohn-Sham density
functional approach and effective field theory.

The plan of the paper is as follows.  
In Sect.~\ref{sect:EFT}, we
review effective field theory for a dilute system of
fermions. 
In Sect.~\ref{sect:effact}, the effective action approach for determining a
Kohn-Sham
energy functional through a Legendre transformation is reviewed.
A systematic approximation procedure for constructing the
energy functional, the inversion method, is presented. 
In Sect.~\ref{sect:results}, the formalism is applied to a dilute Fermi gas in
a harmonic trap and results are presented  through third order (NNLO) in the
dilute EFT expansion using an LDA.  
Section~\ref{sect:summary} summarizes our results and future plans.


\section{EFT for Infinite Dilute Fermi Systems}
\label{sect:EFT}

\subsection{Background}

Effective field theory (EFT) provides a 
powerful framework to study low-energy phenomena
in a model-independent way \cite{LEPAGE89,Birareview,BEANE99}.
The EFT approach is grounded in some very general
physical principles \cite{LEPAGE89}.
If a system is probed or interacts at low energies,
resolution is also low, and fine details of what happens at short
distances or in high-energy intermediate states are not resolved.
Therefore, the short-distance structure can be replaced by something
simpler without distorting the low-energy observables.
This is analogous to a multipole expansion, in which a complicated,
 charge or current distribution is replaced 
for long-wavelength probes by a series
of point multipoles.
EFT uses
local Lagrangian field theory as a framework for carrying out
this program in a complete and systematic way.%
\footnote{Note that conventional nuclear phenomenology also relies on
these principles in using potentials cut off at short distances.
However, cutoff \emph{independence} is a goal of EFT that is not usually
achieved in phenomenological approaches.}
The uncertainty principle implies that high-energy
intermediate states are highly virtual and only last for a short time,
so their effects are not distinguishable from those of local 
operators \cite{LEPAGE89}.
This physics can then be systematically absorbed into the coefficients
of these operators using renormalization.

The effective degrees of freedom (dof's) in an EFT
depend on a separation or resolution
momentum scale $\Lambda$, which sets the radius of convergence of an EFT.
Long-range dof's with respect to $\Lambda$ must be treated explicitly
while short-range physics is encoded in the coefficients of the local
operators. 
(See Ref.~\cite{BEANE99} for details of how the pion can be considered
either a short- or long-distance degree of freedom in the two-nucleon
problem, depending on the resolution scale.) 
The hierarchy of scales in the system is exploited to provide expansion
parameters.
For example, if the typical momenta $k$ are small compared to the 
inverse range of the interaction $1/R$, 
we can take $\Lambda \sim 1/R$ and
low-energy observables can be described by a
controlled expansion in $kR$. All short-distance effects are systematically
absorbed into low-energy constants through renormalization.
As a result,
the EFT approach allows for accurate calculations of low-energy processes 
and properties with well-defined error estimates (based on the
order of truncation) \cite{EFT98,EFT99,BEANE99,Birareview}.

The application of EFT methods to many-body problems promises 
a consistent organization of many-body corrections, with reliable error
estimates, and insight into the analytic structure of observables
(see, for example, the identification using renormalization group
methods of logarithmic contributions to the
energy of dilute systems in Refs.~\cite{BRAATEN97} and \cite{HAMMER00}).
The EFT provides a model-independent description of finite-density
observables in terms of parameters that can be fixed from
scattering in the vacuum or from a subset of finite density properties.
One can also exploit the freedom in an EFT of using different 
regulators and renormalization schemes to find simplifications and
clarifications \cite{HAMMER00}.

While EFT has shown early promise in applications to basic many-body
problems (e.g., Refs.~\cite{BRAATEN97} and \cite{HAMMER00}), there
are formidable challenges in carrying out many-body calculations,
particularly for finite, non-uniform systems.  
For sufficiently small numbers of fermions, 
the many-body Schr\"odinger equation for finite systems
can be solved directly, for example by Green's function Monte Carlo
methods.  However, the computational cost of these methods grows as
a power of the number of particles (or faster), which prevents their
application to very many systems of interest in condensed matter
and quantum chemistry (Coulomb systems) and nuclear physics
(medium to heavy nuclei).

The purpose of the present work is to address these challenges 
systematically. We seek to merge the organizational advantages and
insight provided by EFT  with the calculational power and relative ease
of DFT for finite systems.  
We will build the basic EFT formalism of this merger
for a dilute gas of identical fermions with short-range interactions.
We review below the results from the calculation of the energy 
density in the case of a uniform system~\cite{HAMMER00}. 
These results
will be the starting point of the DFT calculation of the energy density
of a dilute Fermi gas confined by an external harmonic potential.  
They will
also serve as the basis for the LDA to the energy functional.

\subsection{Lagrangian and Energy Density for Uniform System}
\label{dil1a}
We consider
a general local Lagrangian for a nonrelativistic fermion
field that is invariant under Galilean, parity, and time-reversal
transformations:
\bea
  {\cal L}  &=&
       \psi^\dagger \biggl[i\partial_t +\mu + \frac{\nab^{\,2}}{2M}\biggr]
                 \psi - \frac{C_0}{2}(\psi^\dagger \psi)^2
            + \frac{C_2}{16}\Bigl[ (\psi\psi)^\dagger
                                  (\psi\galnab^2\psi)+\mbox{ h.c.}
                             \Bigr]
  \nonumber \\[5pt]
   & & \null +
         \frac{C_2'}{8} (\psi \galnab \psi)^\dagger \cdot
             (\psi\galnab \psi)
+  \ldots\,,
  \label{eq:lag}                                                   
\eea
where $\galnab=\overleftarrow{\nabla}-\nab$ is the Galilean invariant
derivative and h.c.\ denotes the Hermitian conjugate.
The terms proportional to $C_2$ and $C_2'$ contribute to $s$-wave and
$p$-wave scattering, respectively, 
while the dots represent terms with more derivatives and/or more
fields.
The Lagrangian Eq.~(\ref{eq:lag}) represents a particular 
canonical form, which can be reached via field redefinitions.
For example,
higher-order terms with time derivatives are omitted, as they can be
eliminated in favor of terms with spatial derivatives \cite{HAMMER00}.                        

To reproduce the results in Ref.~\cite{HAMMER00},
we can write a conventional generating functional with the Lagrangian
of Eq.~(\ref{eq:lag}) and Grassmann sources coupled to $\psi^\dagger$ and
$\psi$, respectively \cite{NEGELE88}.
The non-quadratic part of the Lagrangian is removed in favor of
functional derivatives with respect to the Grassmann sources and  the
remaining quadratic part is evaluated in terms of a non-interacting
Green's function times the sources.
Perturbative expansions for Green's functions (and subsequently
S--matrix elements) follow by taking
successive functional derivatives, and the ground state energy density
follows by applying the linked cluster theorem (see Ref.~\cite{NEGELE88}
for details).
In calculating the energy, finite density boundary conditions at $T=0$ can be
incorporated into the non-interacting Green's function using the
chemical potential $\mu$ or by including them by hand with a
non-interacting chemical potential.  The latter approach is simplest at
$T=0$ and was adopted in Ref.~\cite{HAMMER00}. 

The coefficients $C_0$, $C_2$, and $C_2'$ can
be obtained from matching the EFT
to a more fundamental theory or to
(at least) three independent pieces of experimental data.
We follow the regularization and renormalization prescription
described in Ref.~\cite{HAMMER00}, namely dimensional regularization with
minimal subtraction, which is particularly convenient for the dilute,
natural system.
By matching to the effective-range expansion for low-energy
fermion-fermion scattering,
we can express the $C_{2i}$ in terms of the effective-range parameters:
\beq
      C_0 = \frac{4\pi a_s}{M},\qquad C_2=C_0 \frac{a_s r_s}{2},
               \quad \mbox{and}\quad
          C_2' = \frac{4\pi a_p^3}{M}
      \label{C2imatch}    \,,
\eeq
where $a_s$ ($a_p$) are the $s$-wave ($p$-wave) scattering length
and $r_s$ is the $s$-wave effective range, respectively.

\begin{figure}[t]
\centerline{\includegraphics*[width=12cm,angle=0]{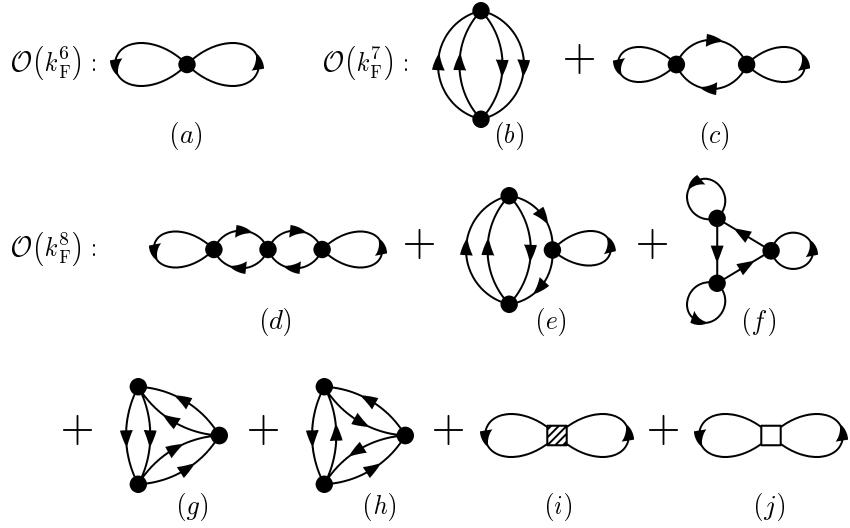}}
\vspace*{-.1in}
\caption{Hugenholtz diagrams for a dilute Fermi gas through order
  $\kf^8$ in the energy density.}
\label{fig:hug}
\end{figure}        

The 
energy density for the uniform dilute fermion system with natural
scattering length \cite{HAMMER00} is calculated as a perturbative
expansion in $\kf/\Lambda$,
where $\kf$ is the Fermi momentum and $\Lambda$ is the resolution
scale (e.g., $\Lambda \approx 1/R$ for hard spheres). 
The non-interacting energy density at zero temperature
for $A$ particles with spin-degeneracy $g$ in volume $V$ can be written as
\beq
\edens_0 =\rho \ {3 \over 5} {\kf^2 \over 2 M} \ ,
\label{eqn:E0}
\eeq
where the density  $\rho$ is
\beq
\rho= \frac{A}{V} =
  g \int {d^3k\over (2\pi)^3} \ \theta (\kf-k) = {g k_F^3\over 6 \pi^2}\ .
\label{unifden}
\eeq
The order-by-order
corrections to Eq.~(\ref{eqn:E0}) due to interactions can be represented
by the Hugenholtz diagrams given in Fig.~\ref{fig:hug}. 
The Feynman rules for calculating
these graphs along with details of the renormalization
needed to render divergent
graphs finite are given in Ref.~\cite{HAMMER00}. 
Here we will only need the final results and we simply quote them
up to next-to-next-to-leading order (NNLO).

The LO diagram of order $\kf^6$ [Fig.~\ref{fig:hug}(a)], 
which represents the Hartree-Fock result,
contributes
\beq
  \edens_1 =\dens\,(g-1)\,\frac{\kf^2}{2M}\,\frac{2}{3\pi}\, \kf a_s
         \ ,
\eeq
where Eqs.~(\ref{C2imatch}) and (\ref{unifden}) have been applied.
At order $\kf^7$ (NLO) there are two diagrams. The three-loop diagram
[Fig.~\ref{fig:hug}(c)]  is an example of an ``anomalous diagram,'' 
which vanishes identically for a uniform system
in the zero-temperature formalism 
but is nonzero when calculated in the
zero-temperature limit. 
In the latter case, its contribution is precisely canceled by the shift
between the noninteracting and interacting chemical potentials,
as dictated 
by the Kohn-Luttinger-Ward theorem \cite{KOHN60,FETTER71,NEGELE88}.
We will find an analogous cancellation in the diagrammatic
expansion of the Kohn-Sham functional.
The other diagram at this order 
[the ``beach ball" diagram of Fig.~\ref{fig:hug}(b)], 
makes the contribution
\beq
   \edens_2 =\dens \,(g-1)\,\frac{\kf^2}{2M}\,(\kf a_s)^2 \frac{4}{35\pi^2}
   \left(11-2\ln2 \right)   \ . 
                \label{energy2d}
\eeq
Finally, we have the graphs of order $\kf^8$ (NNLO). 
The first three are anomalous diagrams.
In addition to graphs
containing the $C_0$ vertex [Figs.~\ref{fig:hug}(g) and (h)],
there also graphs that contain $C_2$ and
$C'_2$. Altogether these graphs give the correction
\bea
      \edens_3 &=&
                \dens\,\frac{\kf^2}{2M}\Bigl[ (g-1)\frac{1}{10\pi}\,(\kf a_s)^2
        \,\kf r_s + (g+1)\,\frac{1}{5\pi}\,(\kf a_p)^3  
                      \nonumber \\
       & &  +(g-1)\{(0.07550\pm 0.00003)+(g-3)\,(0.05741\pm 0.00002)\}\,
           (\kf a_s)^3 \vphantom{\frac{1}{10\pi}} \Bigr] \ ,
       \label{energy3}
\eea
where the integrals for Figs.~\ref{fig:hug}(g) and (h)
have been evaluated numerically.
The expansion can be continued systematically, but this is as far
as we will need here.


\section{Effective Actions and The Inversion Method}
\label{sect:effact}

The energy density given in the last section illustrates that the ground
state energy of a \emph{uniform}, interacting many-body system may be
written as a function of the constant density.
The DFT theorems of Hohenberg and Kohn (HK)~\cite{HK64} 
formally prove the existence of a
generalization to finite, \emph{nonuniform} systems:  
the ground state energy can be
obtained from a functional of the local density alone.
Specifically, for a system with an external potential $v(\xvec)$ coupled
to the density $\rho(\xvec)$, there exists an energy functional
$E[\rho]$ that can be decomposed as
\beq
   E[\rho(\xvec)] = F_{HK}[\rho(\xvec)] 
      + \int\! \dthreex\, v(\xvec)\rho(\xvec) \ ,
      \label{eq:HKuniv}
\eeq
where the functional $F_{HK}[\rho]$, which is known as the HK free energy,
is universal (i.e., independent of the potential $v$).
A variational principle ensures that the functional $E[\rho]$ is
a minimum equal to the ground state energy when evaluated at the exact
ground state density.
The practical problem of density functional theory is to find accurate
and tractable approximations to $F_{HK}[\rho]$
\cite{DREIZLER90,PARR89,ARGAMAN00}.

In applications to Coulomb systems, 
$F_{HK}[\rho]$ is typically decomposed into a noninteracting kinetic
energy $F_{ni}[\rho]$, the Hartree term $E_H[\rho]$, and everything
else, which is defined as the exchange-correlation energy 
$E_{xc}[\rho]$ \cite{ARGAMAN00}:
\beq
   F_{HK}[\rho] = F_{ni}[\rho] + E_H[\rho] + E_{xc}[\rho] \ .
\eeq
In the Thomas-Fermi approximation, the noninteracting functional is evaluated
within an LDA.  This approximation is
accurate at high densities when the characteristic scale over which the
fermion density changes is small compared to the Fermi wavelength.
The approximation is inadequate for most ordinary Coulomb systems and improves
only when some account is taken of the nonlocality in $F_{ni}$.
Kohn and Sham~\cite{KOHN65} introduced a method involving auxiliary
orbits by which $F_{ni}$ could be treated exactly, which has become the
basis for most practical DFT applications \cite{ARGAMAN00,KOHN99}.  

The utility of DFT then rests on finding an explicit
expression, exact or approximate, for $E_{xc}$.  
For Coulomb systems, the conventional procedure is to approximate
$E_{xc}$  in the LDA based on a fit
to Monte Carlo results for the energy of a uniform electron
gas as a function of the density, 
and then to include semi-phenomenological gradient 
corrections \cite{ARGAMAN00}.
For applications to other systems, such as nuclei or trapped atoms, 
we adopt an approximation
scheme based on effective action methods. 
Here we
follow the work of Fukuda {\em et al.} \cite{FUKUDA94,FUKUDA95} 
and its further development 
by Valiev and Fernando \cite{VALIEV97}.
We extend this work by merging it with an effective field theory
expansion.
For consistency with the EFT treatment of dilute systems in
Ref.~\cite{HAMMER00}, we work at zero temperature in Minkowski space.

We begin with the system described by the Lagrangian
in Eq.~(\ref{eq:lag}), to which we add a term for 
an external potential $v(\xvec)$ coupled to the density operator,
$v(\xvec)\psi^\dagger\psi$.%
\footnote{In general, the potential will be coupled to more
 operators than simply $\psi^\dagger\psi$.  
Field redefinitions, which leave observables unchanged, could
be used to eliminate couplings to these operators, which would
be higher order in the EFT expansion \cite{FURNSTAHL01}.
However, such field transformations would also induce energy-dependent
terms in the Lagrangian.
Since we have already used field transformations to achieve a
canonical, energy-independent Lagrangian, we will assume that omitted
higher-order couplings, which should
be suppressed numerically according to power counting, 
can be treated perturbatively.}
Here we take the external potential to
be
an isotropic harmonic confining potential
\beq
v(\xvec)={1\over 2} m\,\omega^2 \, |\xvec|^2 \ ,
\eeq
as might be appropriate for some atomic traps \cite{BRUUN98}, but
the discussion holds for a general external potential.
We also introduce a c-number source, $J(x)$, coupled 
to the composite density operator and write down the generating functional  
using the path
integral formulation,
\beq
    Z[J] = e^{iW[J]}
      = \int\! D\psi D\psi^\dag \ e^{i\int\! d^4x\ [{\cal L}\,+\,   J(x)
    \psi^\dag(x) \psi(x)]}
    \ .
   \label{partfunc1}
\eeq
For simplicity, normalization factors are considered to be implicit
in the functional integration measure.
(See Refs.~\cite{FUKUDA94,FUKUDA95} for a more careful treatment
of the path integrals.)
Using the definition in Eq.~(\ref{partfunc1}),
we see that the density (in the presence of $J$) is
\beq
\rho(x)\equiv \langle \psi^\dag(x)\psi (x)\rangle_J
= {\delta W[J]\over \delta J(x)} \ .
\label{partden1}
\eeq
While it is possible to absorb $v(\xvec)$ into the definition of $J(x)$, 
we find it more convenient to 
recover our original system in the limit
$J\rightarrow 0$.

The effective action is defined
through the functional Legendre transformation 
\beq
\Gamma[\rho]  = 
  W[J]- \int\! d^4x\, 
       J(x)\rho(x)\ .
\label{ehk}
\eeq
This transformation ensures
that $\Gamma$ has no {\em explicit} dependence on
$J$. The existence of such a functional follows from the concavity
of $W[J]$ which guarantees that Eq.~(\ref{partden1}) can be
inverted to give $J=J[\rho]$. The functional dependence  between $J$
and $\rho$ can be used to write Eq.~(\ref{ehk}) entirely in terms of
$\rho$.  
The proof that $W[J]$ is strictly concave is given in
Ref.~\cite{VALIEV97}.

Since we are interested here in describing finite many-body ground states
at $T=0$, it is most convenient to work with functions of the particle
number $A$ rather than the chemical potential $\mu$.
This can be achieved via a conventional Legendre transformation
on $W$ or $\Gamma$, but is more simply carried out implicitly,
by choosing appropriate finite-density boundary conditions that
enforce a given $A$ by hand
(the actual procedure is detailed below).
In the following, we will assume this has been done.
\emph{Thus, $\Gamma$ and $W$ are functions of $A$
and variations over $\rho(x)$ conserve $A$.}
In addition,
we restrict the discussion to time independent sources and densities.  
In this case the effective action acquires a factor that 
corresponds to the time interval over 
which the source is acting, which we indicate schematically as
\beq
\Gamma[\rho]= - E[\rho]\times \int_{-\infty}^{\infty}\, dt \ ,
  \label{eq:LamE}
\eeq
as in Ref.~\cite{FUKUDA94}.
To avoid overly cluttered notation, we will divide out
this ubiquitous time factor everywhere it appears and write
\beq
\til{\Gamma}[\rho]\equiv \Gamma[\rho] 
  \times \left[\int_{-\infty}^{\infty}\, dt\right]^{-1} 
  = - E[\rho]
  \, .
\eeq
and similarly with $W[J]$ and the expansions below.
(We will continue to use $\til{\Gamma}$ rather than $E$ in this section.)

In conventional treatments (e.g., see Refs.~\cite{WEINBERG96,PESKIN95}), 
an effective action is derived from a Legendre
transformation with respect to a source coupled to one of the fields in
the Lagrangian, rather than to a composite operator as in the present
case.
However, the usual advantages of working with an effective action are also
present here.  The effective action has extrema at the
possible quantum ground states of the system, and when evaluated at the
minimum is proportional (at zero temperature) to the ground state
energy \cite{FUKUDA94,FUKUDA95,HU96}.
In particular, Eq.~(\ref{eq:LamE}) defines an energy functional
$E[\rho]$ equal to the ground-state energy when evaluated with the
ground-state density $\rho$. 
 
The extremization condition is shown as follows.
Combining Eq.~(\ref{partden1}) and Eq.~(\ref{ehk}) we find
\beq
   {\delta \til{W}[J]\over\delta J(\xvec)}
    = 
    \int d^3 \yvec\,
    \Biggl({\delta \til{\Gamma}[\rho]\over \delta\rho(\yvec)}\Biggr)
    \Biggl({\delta\rho(\yvec)\over\delta J(\xvec)}\Biggr)
    + \rho(\xvec) + \int d^3 \yvec\,
    \Biggl({\delta \rho(\yvec)\over \delta J(\xvec)}\Biggr) 
    J(\yvec)
\eeq
or
\beq
\int d^3\yvec\ \left({\delta \til{\Gamma}[\rho]\over\rho(\yvec)}
   + J(\yvec)\right)
   \left({\delta\rho(\yvec)\over\delta J(\xvec)}\right)=0
  \ .
\eeq
The invertibility of (\ref{partden1}) implies
\beq
  {\delta\rho(\yvec)\over\delta J(\xvec)}\neq 0 \ ,
\eeq
so we must have
\beq
    {\delta\til{\Gamma}[\rho]\over\delta\rho(\xvec)}
  = -J(\xvec)\ .
\label{min1}
\eeq
The above equation tell us that when $J(\xvec)=0$
the effective action is extremized,
which is a statement of the
second HK theorem \cite{HK64}. 
The strict concavity of $\til{W}[J]$ implies
the strict concavity of $\til{\Gamma}[\rho]$ and so the extremum is a
maximum.
Since $J(\xvec)=0$ corresponds to the original system we see that the 
energy functional is minimized when evaluated at the exact expectation value
of the density.

We observe that the separation of a $v(\xvec)$-dependent part of
the DFT energy functional (or $\til{\Gamma}$) from a universal part
follows directly in the effective action formalism \cite{FUKUDA94}.
From the definitions of $Z$ and $W$, it follows that
\beq
  \til{W}_{v=0}[J] = \til{W}[J+v]
  \label{eq:Wv0}
\eeq
for any $J(\xvec)$.  If we designate $J_\rho(\xvec)$ the inversion of
$\delta\til{W}/\delta J = \rho$ and $J^{0}_\rho(\xvec)$ the inversion
of $\delta\til{W}_{v=0}/\delta J = \rho$ for the same density, 
then 
\beq
  \frac{\delta\til{W}[J_\rho]}{\delta J(\xvec)}
  =   \frac{\delta\til{W}_{v=0}[J^{0}_\rho]}{\delta J(\xvec)}
 =   \frac{\delta\til{W}[J^{0}_\rho+v]}{\delta J(\xvec)} \ ,
\eeq
which implies
\beq
   J_\rho = J^{0}_\rho + v 
    \ .
    \label{eq:Jrho}
\eeq
Upon substituting Eq.~(\ref{eq:Jrho}) into Eq.~(\ref{ehk}), the
effective action becomes
\bea
  \til{\Gamma}[\rho] &=&
      \til{W}[J_\rho] - \int\!\dthreex\, J_\rho(\xvec)\,\rho(\xvec)
    \nonumber \\
    &=& 
    \til{W}_{v=0}[J^{0}_\rho] 
       - \int\!\dthreex\, \Bigl(J^{0}_\rho(\xvec)+v(\xvec)\Bigr)
         \,\rho(\xvec)
    \nonumber \\
    &=& \Bigl[\til{W}_{v=0}[J^{0}_\rho] - \int\!\dthreex\, J^{0}_\rho
    \rho(\xvec)\Bigr] - \int\!\dthreex\,v(\xvec)\,\rho(\xvec)
    \nonumber \\
    &=& \til{\Gamma}_{v=0}[\rho] - \int\!\dthreex\,v(\xvec)\,\rho(\xvec)
     \ ,
\eea
which is the promised result [Eq.~(\ref{eq:HKuniv})]
with an overall minus sign.

Now consider a system that can be characterized by
an effective field theory  power-counting parameter, which
we label $\lambda$.
This parameter may be
dimensionful and can appear to all orders. We will not exhibit it
explicitly here but indicate by subscripts the order in $\lambda$ of a
given function or functional.
In previous discussions of the inversion method 
\cite{FUKUDA94,FUKUDA95}, the parameter $\lambda$ was
always a
coupling constant (e.g., $e^2$ for the Coulomb interaction).
In contrast,
we associate $\lambda$ with an appropriate
EFT expansion parameter.
For example, $\lambda$ could 
be $1/\Lambda$ in the dilute expansion (which we use here) or 
$1/N$ in a large $N$ expansion, where ``$N$'' is the spin degeneracy
$g$ \cite{HAMMER00}.
The effective action functional will depend on $\lambda$ but we
treat $\rho$ and $\lambda$ as independent variables:
\beq
    \til{\Gamma} = \til{\Gamma}[\rho,\lambda]
    \ .
\eeq
However, the ground state expectation value $\rho_g(\xvec)$ will naturally
depend on $\lambda$ as determined by
\beq
   \left.
     \frac{\delta\til{\Gamma}[\rho,\lambda]}{\delta \rho(\xvec)}
    \right|_{{\rho=\rho_g}}
    = 0
     \ .
\eeq
That is, if  $\lambda$ is changed, a different $\rho_g$ will be necessary
to satisfy this equation.

The Legendre transformation defining $\til{\Gamma}$ is
\beq
  \til{\Gamma}[\rho,\lambda] = \til{W}[J,\lambda] - J(\vec1)\rho(\vec1) 
  \ ,
  \label{eq:Gammalam}
\eeq
where $J$ is a functional of $\rho$ and $\lambda$ as well, as dictated by
\beq
   \frac{\delta \til{W}[J,\lambda]}{\delta J(\vec1)} = \rho(\vec1) \ .
\eeq
Here we have introduced the convenient shorthand notation
\beq
   A(\vec1)B(\vec1)\equiv  
  \int \! d^3\, \xvec_1 \, A(\xvec_1)B(\xvec_1)\ .
\eeq
%
%
 As $\lambda$ is changed, $J$  must be adjusted so that  the same
$\rho$ is obtained when taking this derivative. This is how the dependence
of $J$ on $\lambda$ arises;
clearly this dependence can become quite complicated.
 
The inversion method  now proceeds by expanding each of the quantities
that depend on $\lambda$ in \Eq{eq:Gammalam} in
a Taylor series in $\lambda$:
\bea
  J[\rho,\lambda] &=& J_0[\rho] + J_1[\rho] + J_2[\rho] + \cdots \ ,
   \\
  \til{W}[J,\lambda] &=& \til{W}_0[J] + \til{W}_1[J] + \til{W}_2[J] 
  + \cdots \ ,
  \\
  \til{\Gamma}[\rho,\lambda] &=& \til{\Gamma}_0[\rho] + \til{\Gamma}_1[\rho] + \til{\Gamma}_2[\rho] +
  \cdots \ ,
\eea
where, as advertised, the power of $\lambda$ associated with each
function or functional is indicated
by the subscript.
We can substitute the expansion for $J$ into the expansion for
$W$ and do a functional Taylor expansion of $W[J]$ about $J_0$;
this makes the $\lambda$ dependence manifest.
Equating equal powers of $\lambda$ gives ($l=0,1,2,\ldots$)
\bea
  \til{\Gamma}_l[\rho] &=& \til{W}_l[J_0] - J_l(\vec1) \rho(\vec1) +
    \sum_{k=1}^{l} \frac{\delta \til{W}_{l-k}[J_0]}{\delta J_0(\vec1)}
      J_k(1)
      \nonumber \\
      & & \ \qquad\null
      + \sum_{m=2}^{l} \frac{1}{m!}
      \sum_{k_1,\cdots,k_m\geq 1}^{k_1+\cdots+k_m\leq l}
      \frac{\delta^m \til{W}_{l-(k_1+\cdots+k_m)}[J_0]}
       {\delta J_0(\vec1) \cdots \delta J_0(\vec m)} \,
       J_{k_1}(\vec1) \cdots J_{k_m}(\vec m)
       \ .
    \label{eq:Gammalb}
\eea
Since $\rho$ is independent of $\lambda$ \cite{OKUMURA96}, 
each $J_k[\rho]$ follows from each
$\til{\Gamma}_k$:
\beq
  J_k(\vec1)
     = - \frac{\delta \til{\Gamma}_k[\rho]}{\delta \rho(\vec1)}  \ .
\eeq
We  reiterate that all of the $J_l$'s as
defined here are functionals of $\rho$.
 
Starting with the zeroth order expression,
\beq
  \til{\Gamma}_0[\rho] = \til{W}_0[J_0] - J_0(\vec1) \rho(\vec1) \ ,
\eeq
we take its functional derivative with respect to $\rho$:
\beq
 \frac{\delta \til{\Gamma}_0[\rho]}{\delta \rho(\vec1)} = - J_0(\vec1)
 = \frac{\delta \til{W}_0[J_0]}{\delta J_0(\vec1')}\,
   \frac{\delta J_0(\vec1')}{\delta \rho(\vec1)}
   - J_0(\vec1)
   - \rho(\vec1')\frac{\delta J_0(\vec1')}{\delta \rho(\vec1)}
   \ .
   \label{eq:functa}
\eeq
Rearranging,
\beq
  \left(
  \frac{\delta \til{W}_0[J_0]}{\delta J_0(\vec1')} - \rho(\vec1')
  \right)
  \frac{\delta J_0(\vec1')}{\delta \rho(\vec1)}
   = 0 \ ,
\eeq
which implies
\beq
   \rho(\vec1) = \frac{\delta \til{W}_0[J_0]}{\delta J_0(\vec1)} \ ,
   \label{eq:KSq}
\eeq
since the strict concavity of $\til{\Gamma}_0[\rho]$ prohibits
$\delta J_0(\vec1')/\delta \rho(\vec1)$ from having zero eigenvalues.
 
Equation~(\ref{eq:KSq}) says that $J_0(\xvec)$ is the source
(or potential, see below) that
generates the expectation value
$\rho$ from the {\em noninteracting\/} system (that is, the system
defined by $\lambda=0$, which includes the external potential 
and $J_0(\xvec)$ but
no interactions). $J_0(\xvec)$ is not an arbitrary function, 
but the \emph{particular}
one that has this property. 
The existence of a $J_0(\xvec)$ with this property
is the cornerstone of the Kohn-Sham formalism.

Equation~(\ref{eq:KSq}) 
also implies that the second term in \Eq{eq:Gammalb} cancels
with the $k=l$ term of the first sum for all $l>0$, and thus 
$\til{\Gamma}_l$ simplifies to
\bea
  \til{\Gamma}_l[\rho] &=& \til{W}_l[J_0] 
  - \delta_{l,0} J_l(\vec1) \rho(\vec1) +
    \sum_{k=1}^{l-1} \frac{\delta \til{W}_{l-k}[J_0]}{\delta J_0(\vec1)}
      J_k(\vec1)
      \nonumber \\
      & & \qquad\null
      + \sum_{m=2}^{l} \frac{1}{m!}
      \sum_{k_1,\cdots,k_m\geq 1}^{k_1+\cdots+k_m\leq l}
      \frac{\delta^m \til{W}_{l-(k_1+\cdots+k_m)}[J_0]}
       {\delta J_0(\vec1) \cdots \delta J_0(\vec m)} \,
       J_{k_1}(\vec1) \cdots J_{k_m}(\vec m)
       \ .
    \label{eq:Gammal}
\eea
These equations allow us to build the $\til{\Gamma}_l$'s recursively.
Note that the $\til{W}_k$ functionals have the same diagrammatic
expansion as in Fig.~\ref{fig:hug}, but the fermion lines are evaluated
with Kohn-Sham (KS) propagators (see below).
For a given $l$, we only need $\til{W}_k$'s with $k$ less than or equal to
 $l$ and $J_k$'s with $k$ smaller than $l$ 
 (which means lower-order $\til{\Gamma}_k$'s).
We will illustrate the procedure by constructing  the first few orders.

Since the lowest-order term in $\til{W}[J]$ corresponds to the system
without interactions between the fermions, 
we can write $\til{W}_0[J]$ explicitly by introducing normalized
single-particle orbitals that satisfy the equation
\beq
 \left( -{\nabla^2 \over 2M} + v(\xvec) - J_0(\xvec) \right)
   \varphi_i (\xvec) 
   = \varepsilon_i \varphi_i (\xvec) \ .
 \label{eq:ks1}  
\eeq
The index $i$ represents all quantum numbers except for the spin (we
consider only spin-independent interactions here).
$\til{W}_0[J_0]$ is then (minus) the sum of the single-particle
eigenvalues up to the Fermi energy $\varepsilon_{\rm F}$
(which is equal to the chemical potential)
\beq
\til{W}_0[J_0]=  -g \sum_{\varepsilon_i <\varepsilon_{\rm F}} 
   \varepsilon_i\ .
   \label{eq:Wsum}
\eeq
[Equation~(\ref{eq:Wsum}) can be derived by evaluating
$W_0[J_0] \propto \Tr\ln(\greenKS^0)^{-1}$ using the Kohn-Sham
Green's function $\greenKS^0$ defined below in Eq.~(\ref{eq:Gks}).]
In practice, $\varepsilon_{\rm F}$ is determined by simply counting
orbitals until the $A$ lowest are filled (accounting for the spin
degeneracy $g$).
Note that the $\varepsilon_i$'s are functionals of $J_0$ through 
Eq.~(\ref{eq:ks1}); using the normalization of the $\varphi_i$'s, we find
\beq
   \frac{\delta \varepsilon_i}{\delta J_0(\yvec)}
     = \frac{\delta}{\delta J_0(\yvec)}
       \int\! \dthreex\, \varphi_i^*(\xvec)
     \left( -{\nabla^2 \over 2M} + v(\xvec) - J_0(\xvec) \right)
         \varphi_i (\xvec) 
    = -\varphi_i^*(\yvec)\varphi_i(\yvec) \ . 
\eeq
Equations~(\ref{eq:KSq}) and (\ref{eq:Wsum}) 
show that the density may be written as
\beq
\rho(\xvec) =  
     - g\sum_i^{\rm occ.}\, 
           \frac{\delta \varepsilon_i}{\delta J_0(\yvec)}
    =  g\sum_i^{\rm occ.}\, 
     \varphi^*_i (\xvec) \varphi_i (\xvec) \ ,
     \label{eq:famous}
\eeq
where the sum is over occupied (``occ.'') states.
Equation~(\ref{eq:famous}) corresponds to
the famous result of Kohn and Sham,
 which gives the {\em exact} ground state
density in terms of the orbitals of a non-interacting system
\cite{KOHN65}. 

With the above results, the lowest order effective action is
\beq
\til{\Gamma}_0[\rho]=
  -g\sum_i^{\rm occ.} \varepsilon_i 
    - \int \! \dthreex \ J_0 (\xvec)\ \rho(\xvec)
  \ .
\label{Gam0}
\eeq
We can also use Eq.~(\ref{eq:ks1}) to eliminate $\varepsilon_i$ from 
Eq.~(\ref{Gam0}) so that it reads
\beq
\til{\Gamma}_0[\rho]=-T_s[\rho]-\int \! \dthreex \ v(\xvec)\ \rho(\xvec) \ ,
\label{G02}
\eeq
where 
\beq
 T_s[\rho]= g\sum_i^{\rm occ.}\int \! \dthreex \  
 \varphi_i^* (\xvec)\left( -{\nabla^2 \over 2M}\right) \varphi_i (\xvec) 
 \label{T0}
 \eeq
is the total kinetic energy of the KS non-interacting system.
If $\til{\Gamma}_0[\rho]$ were 
given as an explicit functional of  $\rho$, then $J_0[\rho]$ could be
determined by taking a functional derivative according to
Eq.~(\ref{eq:functa}). However, taking the functional
derivative of the expression in Eq.~(\ref{Gam0}) merely reproduces 
the result of Eq.~(\ref{eq:KSq}).
Instead, we follow Ref.~\cite{VALIEV97} and determine $J_0$ from
the \emph{interacting} effective action, which we now construct.

From Eq.~(\ref{eq:Gammal}) we can find $\til{\Gamma}_1[\rho]$ since
\beq
   \til{\Gamma}_1[\rho] = \til{W}_1[J_0[\rho]] \ . 
   \label{eq:G1j}
\eeq
For the dilute Fermi system, 
this is easily calculated. 
We first introduce the Green's
function of the KS non-interacting system, $\greenKS^0 (x,x')$, 
which satisfies 
\beq
\left( i \partial_t+ {\nabla^2 \over 2M}-v(\xvec)+J_0(\xvec)\right) 
\greenKS^0 (\xvec t,\xvec't')=\delta^3(\xvec-\xvec')\delta(t-t')
\label{eq:GKSEq}
\eeq
with finite density boundary conditions \cite{FETTER71}.
The KS Green's function has the usual spectral decomposition in terms
of the orbitals of Eq.~(\ref{eq:ks1}):
\beq
   i\greenKS^0 (\xvec t,\xvec't')=\sum_i \varphi_i (\xvec)\,
       \varphi_i^* (\xvec')\,  
   e^{-i\varepsilon_i(t-t')}[\theta(t-t')\, 
           \theta(\varepsilon_i-\varepsilon_{\rm F})
   -\theta(t'-t)\, \theta(\varepsilon_{\rm F}-\varepsilon_i)]\ .
  \label{eq:Gks}
\eeq
The Feynman rules in position space follow
conventionally from Eq.~(\ref{eq:lag}) 
 \cite{NEGELE88,HAMMER00} and we have from 
Fig.~\ref{fig:hug}(a)
[with fermion lines representing $i\greenKS^0$]
\beq
\til{W}_1[J_0]= {1\over2}\, g\,(g-1)\, C_0 \int \dthreex\ 
    \greenKS^0 (x,x^+)
  \, \greenKS^0 (x,x^+)  \ ,
\label{eq:W1}
\eeq 
where the right side is independent of $x_0$ by Eq.~(\ref{eq:Gks}).
The Green's function with equal arguments can be directly expressed in
terms of the density,
\beq
  \rho(\xvec) = -ig\,\greenKS^0(x,x^+)  \ .
  \label{eq:denG}
\eeq 
Using this result and Eq.~(\ref{eq:G1j}),
we have
\beq 
\til{\Gamma}_1[\rho]= -{1\over2} {(g-1)\over g} \ C_0 \int \dthreex 
   \ |\rho (\xvec)|^2 \ .
\label{eq:G1r}
\eeq
Since the dependence on $\rho(\xvec)$ is explicit
in  Eq.~(\ref{eq:G1r}), we can directly take
the functional derivative with respect to  
$\rho$ to obtain
\beq
J_1(\xvec)=  {C_0\,(g-1) \over g } \ \rho (\xvec)\ .
\label{eq:J1r}
\eeq

Direct functional
derivatives with respect to $\rho$ will not be possible at
higher order.  
However, we can find functional derivatives with respect to
$J_0$.
An alternative path to $J_1[\rho]$ from
$\til{\Gamma}_1[\rho]$ is
\beq
  J_1(\xvec) = - \frac{\delta \til{\Gamma}_1[\rho]}{\delta \rho(\xvec)}
    = - \int\!\dthreey\,
      \frac{\delta \til{\Gamma}_1[\rho]}{\delta J_0(\yvec)} \,
    \frac{\delta J_0(\yvec)}{\delta \rho(\xvec)}
    = \int\!\dthreey\,  \Dfunct(\xvec,\yvec) \, 
       \frac{\delta \til{W}_1[J_0]}{\delta J_0(\yvec)}  \ ,
    \label{eq:J1}
\eeq
which defines the inverse ``density-density'' correlator
\beq
  \Dfunct(\xvec,\yvec) \equiv  
     -\frac{\delta J_0(\yvec)}{\delta \rho(\xvec)}
    = -\left(
    \frac{\delta^2 \til{W}_0[J_0]}{\delta J_0(\xvec)\,\delta J_0(\yvec)}
    \right)^{-1} \ .
\eeq
We can find an expression for $\Dfunct$
by taking the functional derivative with respect to $J_0$ in
Eq.~(\ref{eq:W1}). Remembering that $\greenKS^0$ is a functional of
$J_0$ it is straightforward to show that 
\beq
{\delta \greenKS^0 (x_1,x_2) \over \delta J_0(\xvec) }= 
-\int \greenKS^0 (x_1,x) \greenKS^0 (x,x_2)\ dx_0\ .
\label{Gderiv}
\eeq
Using Eq.~(\ref{Gderiv}), the derivative of $\til{W}_1$ becomes
\bea
{\delta \til{W}_1[J] \over \delta J_0(\xvec)}&=&
  - g(g-1)\ C_0  \int\! d^4 y\, dx_0 \ \greenKS^0 (y,x) \greenKS^0 (x,y^+)
   \greenKS^0 (y,y^+)\nonumber \\
   &=& -i (g-1)\ C_0  \int\! d^4 y\, dx_0 \ \greenKS^0 (y,x)\,
   \greenKS^0 (x,y^+)\, \rho(\yvec)\ .
  \label{W1deriv}
 \eea
Comparing Eq.~(\ref{W1deriv}) to Eq.~(\ref{eq:J1}) we find
\beq
\Dfunct(\xvec,\yvec)= 
{i \over g} \left[\int dy_0\ \greenKS^0 (x,y) \greenKS^0(y,x)\right]^{-1}\, .
\label{eq:Dfunct}
\eeq
This correlator will appear in all higher-order contributions.

\begin{figure}[t]
\centerline{\includegraphics*[width=6.0in,angle=0]{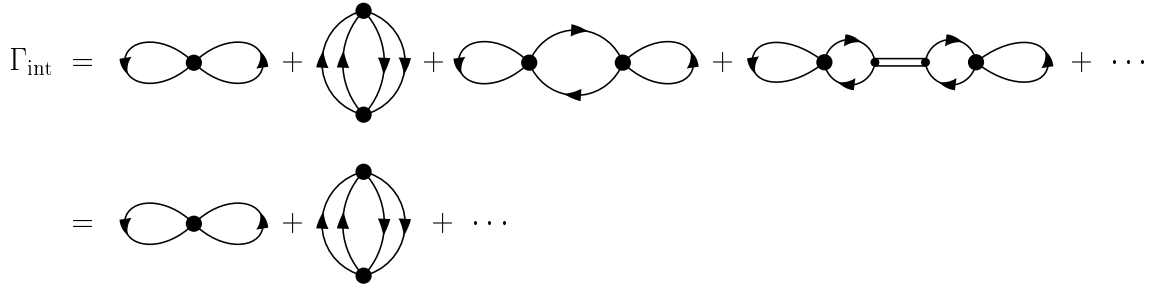}}
\vspace*{-.1in}
\caption{Hugenholtz diagrams for the LO and NLO contributions
    to the Kohn-Sham
   interaction effective action $\Gamma_{\rm int}$.
   The cancellation of the last two diagrams on the first line
   is given by Eq.~(\ref{eq:cancel}).}
\label{fig:Gamma_int}
\end{figure}        

Having determined $J_1[\rho]$, we can find $\Gamma_2[\rho]$ from
  \bea
   \til{\Gamma}_2[\rho] &=& \til{W}_2[J_0]
   + \int\!\dthreex\,
   \frac{\delta \til{W}_1[J_0]}{\delta J_0(\xvec)} J_1(\xvec)
   + \frac{1}{2}\int\!\dthreex\,\dthreey\,
      \frac{\delta^2 \til{W}_0[J_0]}{\delta J_0(\xvec)\,\delta J_0(\yvec)}
      J_1(\xvec)\, J_1(\yvec)
      \nonumber \\[5pt]
      &=&  \til{W}_2[J_0]
   + \frac{1}{2}\int\!\dthreex\,\dthreey\,
     \frac{\delta \til{W}_1[J_0]}{\delta J_0(\xvec)}
     \Dfunct(\xvec,\yvec)
     \frac{\delta \til{W}_1[J_0]}{\delta J_0(\yvec)}\ .
     \label{eq:G2}
  \eea
$\til{W}_2[J_0]$ is calculated from the graphs Figs.~\ref{fig:hug}(b)
and (c):
\bea
\til{W}_2[J_0]&=& ig (g-1){C_0^2 \over 4} \ \int\! d^4x\  d^4 y\  
 \greenKS^0 (x,y)\greenKS^0 (x,y)
 \greenKS^0(y,x)\greenKS^0(y,x)\nonumber \\
&& \hspace*{3mm} \null - ig(g-1)^2\ {C_0^2 \over 2} \int\! d^4x \ d^4 y \ 
\greenKS^0 (x,x^+)\greenKS^0 (x,y) \greenKS^0(y,x)\greenKS^0(y,y^+)\ .
\label{eq:W2}
\eea
By using Eqs.~(\ref{W1deriv}) and (\ref{eq:Dfunct}),
 we can show explicitly
that the second term in the expression for $\til{\Gamma}_2$ 
exactly cancels the second term
in $\til{W}_2$. First note that
\begin{figure}[t]
\centerline{\includegraphics*[width=7.2cm,angle=0]{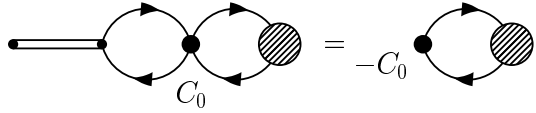}}
\vspace*{-.1in}
\caption{General cancellation of the inverse 
  correlator $\Dfunct$ for zero-range interactions with no
  derivatives.}
\label{fig:vxc}
\end{figure}        
\bea
   \int\!\dthreey\,\Dfunct(\xvec,\yvec)\, 
  \frac{\delta \til{W}_1[J_0]}{\delta J_0(\yvec)}&=&
   -iC_0\, (g-1) \int\!\dthreey\, \left[\int\! dy_0\ 
   \greenKS^0 (x,y) \greenKS^0(y,x)\right]^{-1}
 \nonumber \\
 & & \null \times
 \int \! d^4 w\! \int \! dy_0\,\greenKS^0 (w,y) \greenKS^0 (y,w^+)
   \greenKS^0 (w,w^+)
 \nonumber \\
 &=& -i C_0\, (g-1)\, \greenKS (x,x^+)\, .
\eea
We now see that
\bea
 \lefteqn{{1\over2}\int\!\dthreex\,\dthreey\, 
   \frac{\delta \til{W}_1[J_0]}{\delta J_0(\xvec)}\,
     \Dfunct(\xvec,\yvec)\, 
     \frac{\delta \til{W}_1[J_0]}{\delta J_0(\yvec)}}
   \nonumber \\ & &  
    \qquad\qquad  =
     ig(g-1)^2\ {C_0^2 \over 2} \int\! d^4x \ d^4 y \ 
     \greenKS^0 (x,x^+)\greenKS^0 (x,y) \greenKS^0(y,x)\greenKS^0(y,y^+)\, ,
     \label{eq:cancel}
\eea
which is the negative of the second term in Eq.~(\ref{eq:W2}). 
This cancellation is shown in Fig.~\ref{fig:Gamma_int}, where $\Dfunct$
is represented by a double line.
Thus, Eq.~(\ref{eq:G2}) reduces to
\beq
\til{\Gamma}_2[\rho]= ig (g-1)\,{C_0^2 \over 4} \int\! d^4x\  d^4 y\  
\greenKS^0 (x,y)\greenKS^0 (x,y) \greenKS^0(y,x)\greenKS^0(y,x) \, .
  \label{eq:tilgam2}
\eeq

The
second term in $W_2$ corresponds to  the \lq\lq anomalous" graph (c) in
Fig.~\ref{fig:hug}. 
This graph vanishes identically in the uniform system at zero
temperature. However, it does not vanish at finite temperature or in a
finite system (at any temperature).   Rather, the 
cancellation exhibited above
is analogous to the Kohn-Luttinger-Ward theorem mentioned earlier.
Similar cancellations, of the type illustrated diagrammatically in
Fig.~\ref{fig:vxc}, completely eliminate contributions of $\Dfunct$ to the
effective action up to N${}^3$LO in the EFT expansion for short-range
forces.
This complete
cancellation does \emph{not} occur for long-range forces, or if the
zero-range delta functions at the $C_0$ vertices
are regulated by a cutoff rather than by
dimensional regularization, as used here.

\begin{figure}[t]
\centerline{\includegraphics*[width=12cm,angle=0]{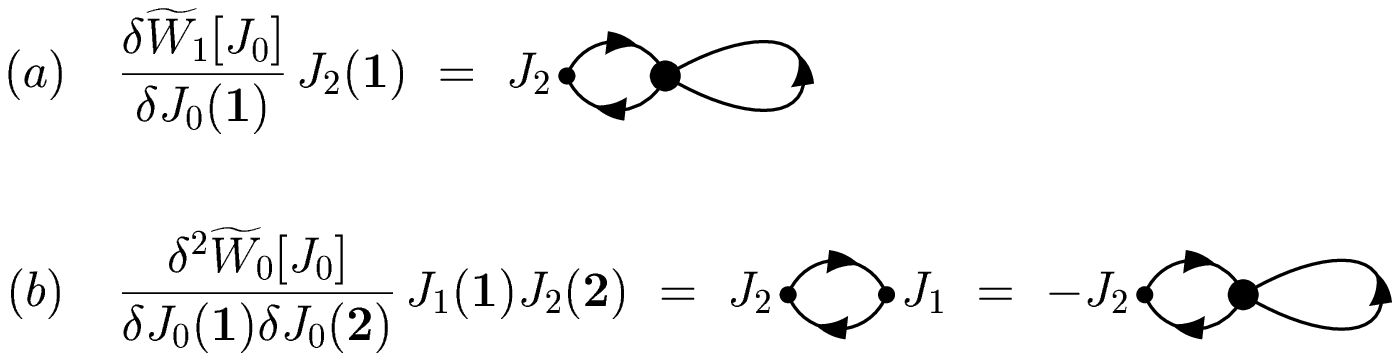}}
\vspace*{-.1in}
\caption{Cancellation of contributions to $\til{\Gamma_3}$ involving
  $J_2$ [see Eq.~(\ref{eq:Gamma3})].  The graphical representation
  of $J_1$ (with an important minus sign)
  comes from Eqs.~(\ref{eq:J1r}) and (\ref{eq:denG}).}
\label{fig:J2cancel}
\end{figure}

\begin{figure}[t]
\centerline{\includegraphics*[width=12cm,angle=0]{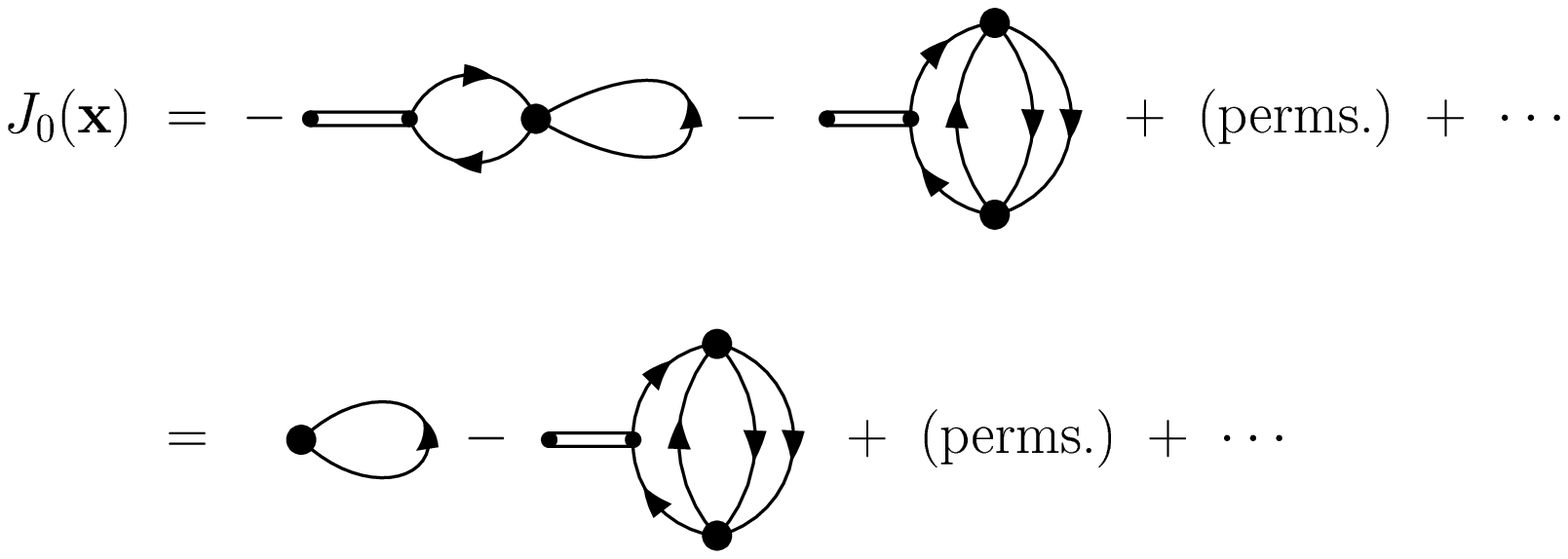}}
\vspace*{-.1in}
\caption{Contributions to the Kohn-Sham potential $J_0$
 through NLO [see Eq.~(\ref{eq:final})].}
\label{fig:J0}
\end{figure}

We will illustrate the construction of $\til{\Gamma}_3$ graphically to
indicate how the cancellations occur at NNLO.
We start with the explicit functional expression:
\bea
  \til{\Gamma}_3 &=& \til{W}_3[J_0] + 
    \frac{\delta \til{W}_2[J_0]}{\delta J_0(\vec1)}J_1(\vec1)
    + \frac{\delta \til{W}_1[J_0]}{\delta J_0(\vec1)}J_2(\vec1)
    + \frac12\, 
       \frac{\delta^2\til{W}_1[J_0]}{\delta J_0(\vec1)\,\delta J_0(\vec2)}
       J_1(\vec1) J_1(\vec2)
    \nonumber \\
    & & \null 
    +  \frac{\delta^2\til{W}_0[J_0]}{\delta J_0(\vec1)\,\delta J_0(\vec2)}
       J_1(\vec1) J_2(\vec2)
    + \frac16\, 
       \frac{\delta^3\til{W}_1[J_0]}{\delta J_0(\vec1)\,\delta J_0(\vec2)\,
          \delta J_0(\vec3)}
       J_1(\vec1) J_1(\vec2) J_1(\vec3)
    \ .
    \label{eq:Gamma3}
\eea
Figure~\ref{fig:J2cancel} illustrates 
that the two terms with $J_2$ factors cancel with
each other. 
Every functional differentiation with respect to $J_0$, which inserts
a density operator, is represented by a small dot (the large dot
is a $C_0$ vertex).
$\til{W_1}$ corresponds to Fig.~\ref{fig:hug}(a), so the
functional derivative in Fig.~\ref{fig:J2cancel}(a) yields the
diagram shown; an explicit diagram for $J_2$ is not needed.
Similarly, the second term is represented in Fig.~\ref{fig:J2cancel}(b).
As shown, substituting $J_1$ (see Fig.~\ref{fig:J0})
yields minus the first contribution,
so the sum is zero and $J_2$ does not appear in
$\til{\Gamma}_3$.

\begin{figure}[t]
\centerline{\includegraphics*[width=7.5cm,angle=0]{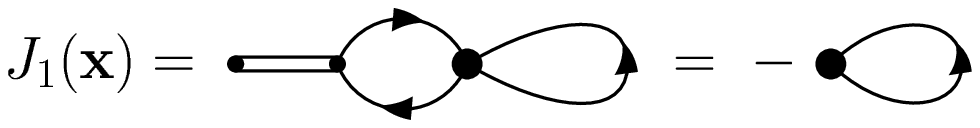}}
\vspace*{-.1in}
\caption{Graphical representation of the Kohn-Sham potential $J_1(\xvec)$
   from Eqs.~(\ref{eq:J1r}) and (\ref{eq:J1}).}
\label{fig:Jone}
\end{figure}        
\begin{figure}[t]
\centerline{\includegraphics*[width=14.5cm,angle=0]{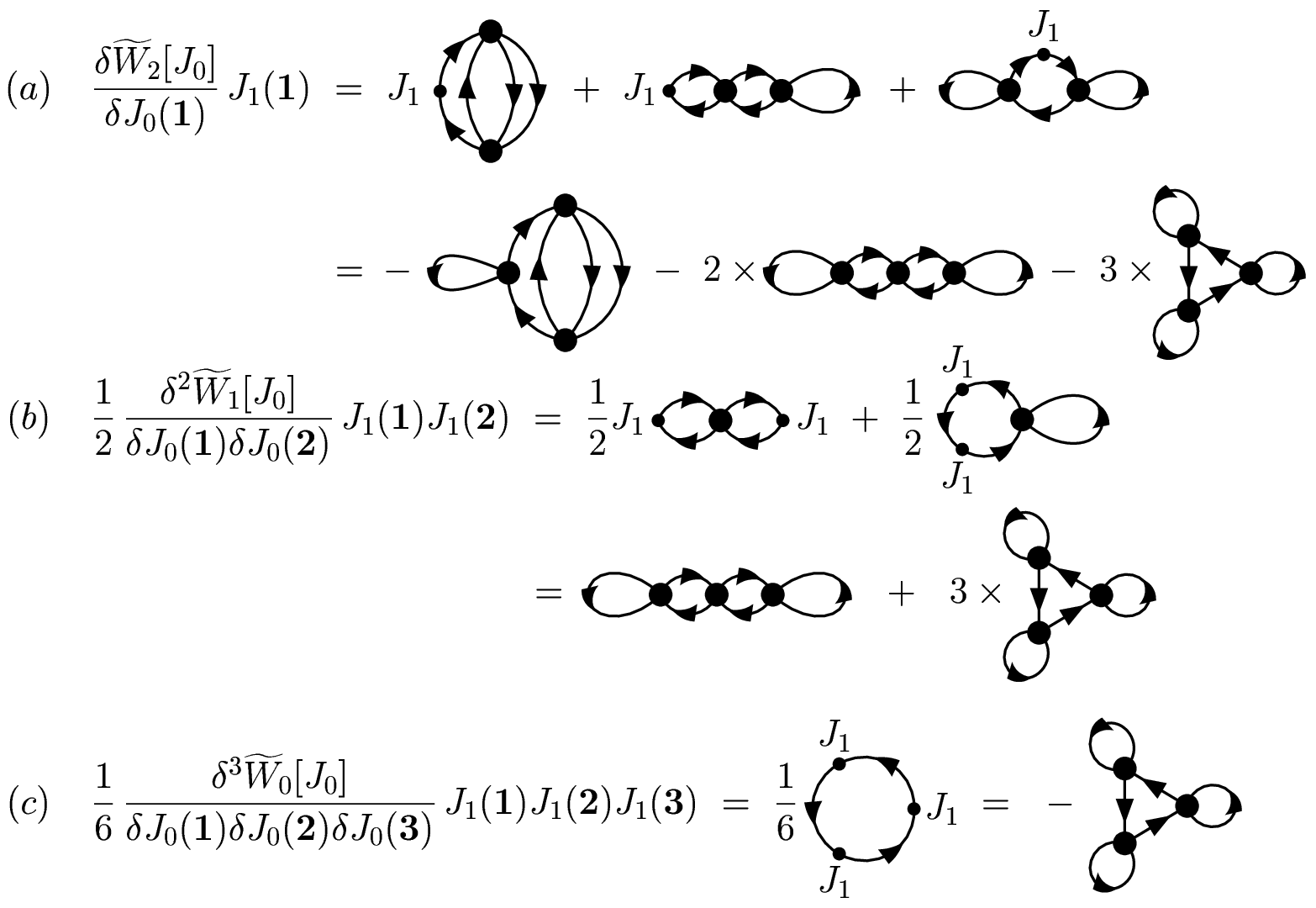}}
\vspace*{-.1in}
\caption{Cancellation of contributions to $\til{\Gamma_3}$ with
  anomalous diagrams in Fig.~\ref{fig:hug}. The graphical representation
  of $J_1$ (with an important minus sign)
  comes from Eqs.~(\ref{eq:J1r}) and (\ref{eq:denG}).}
\label{fig:Gam3cancel}
\end{figure}        

Thus, $\til{\Gamma}_3$ reduces to
\bea
  \til{\Gamma}_3 &=& \til{W}_3[J_0] + 
    \frac{\delta \til{W}_2[J_0]}{\delta J_0(\vec1)}J_1(\vec1)
    + \frac12\, 
       \frac{\delta^2\til{W}_1[J_0]}{\delta J_0(\vec1)\,\delta J_0(\vec2)}
       J_1(\vec1) J_1(\vec2)
    \nonumber \\
    & & \null 
    + \frac16\, 
       \frac{\delta^3\til{W}_1[J_0]}{\delta J_0(\vec1)\,\delta J_0(\vec2)\,
          \delta J_0(\vec3)}
       J_1(\vec1) J_1(\vec2) J_1(\vec3)
    \ .
    \label{eq:Gamma3b}
\eea
$\til{W}_3$ is given by the sum of Hugenholtz graphs in
Fig.~\ref{fig:hug}(d) through (j).  
But the sum of the other terms in Eq.~(\ref{eq:Gamma3b}) exactly
cancel the diagrams in Fig.~\ref{fig:hug}(d), (e), and (f), as
shown in Fig.~\ref{fig:Gam3cancel}.
The factors in front of the Feynman diagrams indicate additional
multiplicative factors beyond those prescribed by the Feynman rules.
These factors conspire to precisely subtract the anomalous diagrams
in $\til{W}_3$, leaving only Fig.~\ref{fig:hug}(g) through (j).

All higher orders in $\til{\Gamma}[\rho,\lambda]$ are determined in a similar
manner.
Direct calculation becomes cumbersome, but one can formulate
Feynman rules for $\til{\Gamma}$, which dictate how to make 
appropriate insertions of $\Dfunct$;
these are given in Refs.~\cite{FUKUDA95,YOKOJIMA95,OKUMURA96,VALIEV97}. 
It is important to 
note that the Kohn-Sham potential $J_0$ completely determines
each order in the expansion of $\til{\Gamma}$.

The cancellations exhibited here at low orders are expected from a
comparison of the DFT/EFT expansion to perturbation theory.  
In perturbation theory for the confined system,
the energy is calculated 
by evaluating the diagrams in Fig.~\ref{fig:hug} using  
the noninteracting propagator in the presence of the external potential
for the fermion lines.
(This propagator
would take the form of Eq.~(\ref{eq:Gks}) with harmonic oscillator
wave functions for the orbitals.)
Using the self-consistent $\greenKS$ instead sums an infinite class of
higher-order diagrams at each order.
The cancellation of anomalous diagrams from $\til{\Gamma}$ corresponds to
the removal of contributions already included through self-consistency
(e.g., tadpoles). 
Although the nonperturbative contributions for the dilute, natural
system are not required by power counting, the self-consistent
calculation of the energy and density together is actually easier in
practice than the purely perturbative calculation.

The appearance of the inverse density-density correlator $\Dfunct$ can
be understood by comparison to the effective actions for local fields
and non-local composite fields.  
In the former case, the Legendre transformation removes
one-particle intermediate states
(leaving only one-particle-irreducible diagrams), 
while in the latter case, the Legendre
transformation removes two-particle intermediate states
(leaving two-particle-irreducible diagrams).
Thus we infer that the role of $\Dfunct$ is to remove intermediate states
created by $\psi^\dagger\psi$.  The difference here is that we cannot
write a closed-form expression for the effective action, which is
possible in the other cases.   However, we have seen that for
short-range interactions, the extra diagrams at low orders cancel against
anomalous diagrams, which is a great simplification.

To find an expression for $J_0$, 
we apply the variational principle satisfied by
$\til{\Gamma}[\rho,\lambda]$ to its expansion:
\bea
   \left.\frac{\delta \til{\Gamma}[\rho,\lambda]}{\delta \rho(\vec1)}
  \right|_{\rho=\rho_{\rm gs}} &=& 0
  \nonumber \\
  &=&
  \left.\frac{\delta (\til{\Gamma}_0[\rho] +\Gamint[\rho])}{\delta \rho(\vec1)}
  \right|_{\rho=\rho_{\rm gs}}
  =
  -J_0(\vec1)|_{\rho_{\rm gs}} +
   \left.\frac{\delta \Gamint[\rho]}{\delta \rho(\vec1)}
  \right|_{\rho=\rho_{\rm gs}}
  \ ,
\eea
where the interaction effective action is
\beq
  \Gamint[\rho] \equiv \sum_{i=1} \til{\Gamma}_i[\rho] \ ,
\eeq
or
\beq
  J_0(\vec1)\Bigr|_{\rho=\rho_{\rm gs}} = 
    \left.\frac{\delta \Gamint[\rho]}{\delta \rho(\vec1)}
  \right|_{\rho=\rho_{\rm gs}}
 =
 - \Dfunct(\vec1,\vec1') \,
    \left.\frac{\delta \Gamint[\rho]}{\delta J_0(\vec1')}
  \right|_{\rho=\rho_{\rm gs}}
    \ .
    \label{eq:final}
\eeq
We stress that these relations hold only
when we are solving for the Kohn-Sham potential corresponding
to the ground-state density $\rho_{\rm gs}(\xvec)$.

The second equality in Eq.~(\ref{eq:final}), 
which is shown diagrammatically through NLO in
Fig.~\ref{fig:J0}, is the key to the general
Kohn-Sham self-consistent procedure:
\begin{enumerate}
 \item Choose an approximation for $\Gamint[\rho]$ by
 truncating the expansion in $\lambda$ at some order.
 \item Make a reasonable initial guess for the Kohn-Sham potential
 $J_0(\xvec)$.
 \item Calculate $\Gamint[\rho]$ starting from $J_0$.
 \item Use \Eq{eq:final} to determine a new Kohn-Sham
 potential $J_0(\xvec)$.
 \item Repeat the last two steps until self-consistency is reached
 (i.e., until some measure of the change in $J_0$ is less than
 a given tolerance).
\end{enumerate}
In the next section, we will apply an LDA to $\til{\Gamma_{\rm int}}$,
which means that it will given explicitly as a functional of $\rho$.
In this case, the second equality in Eq.~(\ref{eq:final}) is superfluous. 
One can also avoid explicitly evaluating
Eq.~(\ref{eq:final}) by adjusting
$J_0(\xvec)$ using a steepest-descent approach \cite{VALIEV97}.

To this point we have neglected to mention that expressions such
as $\til{\Gamma}_2$ in Eq.~(\ref{eq:tilgam2})
are divergent.
For a uniform system, $J_0$ is constant.
In this case,  $\til{W}_2$ 
and subsequently $\til{\Gamma}_2$ are renormalized by
dimensional regularization and minimal subtraction as in 
Ref.~\cite{HAMMER00}.
In a finite system, with $J_0$ a function of $\xvec$,
the ultraviolet linear divergence in Eq.~(\ref{eq:tilgam2})
is renormalized by the same counterterm \cite{COLLINS86},
but it is computationally awkward to renormalize in the finite system.
By using a derivative expansion, we can perform all renormalizations
in the uniform system.
This will be carried out explicitly in future work.
With the LDA truncation applied below,
we can simply use the renormalized expressions for the
energy density from Ref.~\cite{HAMMER00} to calculate the renormalized
Kohn-Sham potential and energy functional.
 
\section{Results for Dilute Fermi System in a Trap}
\label{sect:results}

In this section, we present representative numerical
results for the dilute Fermi system defined
in Sect.~\ref{sect:EFT} when confined in a harmonic oscillator trap.
Our principal goal here is to illustrate the nature of the convergence
of the EFT in a finite system both qualitatively and quantitatively.
We consider effective range parameters corresponding to both
attractive and repulsive underlying interactions (but we do not allow
for pairing).
To emphasize the difference between Kohn-Sham (KS) and Thomas-Fermi (TF)
approaches and
since we are ultimately interested in nuclear systems, we consider
relatively small numbers of trapped fermions.
Current experiments with trapped atoms use $10^5$--$10^6$ atoms
\cite{GRANADE02};
as noted below, 
for such large systems the differences between KS and TF get washed out.

\subsection{The Local Density Approximation (LDA)}

To carry out the Kohn-Sham self-consistent procedure, we need to
evaluate the expressions in the expansion of $\til{\Gamma}[\rho]$,
so that we can use Eq.~(\ref{eq:final}) to find $J_0(\xvec)$.
In applications to Coulomb systems, 
the Kohn-Sham energy functional is conventionally written as
\beq
E[\rho]= T_s[\rho] + \int\! \dthreex \, v(\xvec)\rho(\xvec)
   +E_{\rm H}[\rho]+E_{xc}[\rho] \ ,
\label{Efunc}
\eeq
where $T_s$ is given in Eq.~(\ref{T0}), $E_{\rm H}$ is the Hartree
energy and $E_{xc}[\rho]$ is known as the exchange-correlation energy.
The Hartree energy is singled out because it can be explicitly written
in terms of the density.
When we have contact interactions only, the Fock term has the same
dependence on the density as the Hartree term, and so we can also include it
explicitly and replace $E_{\rm H}$ by $E_{\rm HF}$, redefining
$E_{xc}$ appropriately.

This decomposition corresponds to writing the effective action 
as \cite{VALIEV97}
\beq
 \til{\Gamma}[\rho]= \til{\Gamma}_0[\rho]+\til{\Gamma}_1[\rho]+
\sum_{i=2}^{\infty}\,  \til{\Gamma}_i[\rho] \ ,
\eeq
up to an overall minus sign.
The expression for $\til{\Gamma}_0$ given in Eq.~(\ref{G02}) shows that
it has the same form as the first two terms in the energy functional in
Eq.~(\ref{Efunc}). In general, the explicit functional dependence of
$T_s$ on the density is unknown since it enters implicitly through the
orbitals.
In practice it is easiest to calculate the  first two terms by
writing them as in \Eq{Gam0}:
\beq
\til{\Gamma}_0[\rho]
  =  -g\sum_i \varepsilon_i - \int \! \dthreex \ J_0 (\xvec)\ \rho(\xvec)
\label{Gam0b}  \ .
\eeq
The other terms in $\til{\Gamma}$ may be written using the Green's
function for the KS non-interacting system
as shown in Sec.~\ref{sect:effact}. The KS Green's functions
are explicit functionals of  $J_0$ {\em not} $\rho$, however, and therefore
almost all of $\til{\Gamma}$ is not given as an explicit functional of
the density;
the general exception is the Hartree term and here the
Fock term since we have only contact interactions. 
Furthermore, the actual expressions are quite difficult to evaluate in a
finite system.

As a first approximation, we use the LDA, which 
may be considered the lowest-order term in a derivative
expansion of the energy functional. The idea is to
expand around the uniform system where (as was shown in Section
\ref{dil1a}) the energy functional can be written as an explicit
function of $\rho$. 
The LDA prescription is
\beq
E_{xc}^{\rm LDA}[\rho(\xvec)]\equiv 
\int \! \dthreex\,\edens_{2+}(\rho_0)|_{\rho_0\rightarrow \rho(\xvec)}\, ,
\eeq
where $\edens_{2+}$ is the EFT energy density of the uniform system,
including terms at second order and higher.

By combining results from Secs.~\ref{dil1a} and \ref{sect:effact},
we may write the exchange-correlation energy, to
third order (NNLO) in the EFT expansion, as
\bea
  E^{\rm LDA}_{xc}[\rho(\xvec)]
    &=& \int \! \dthreex\,  \left\{ \edens_2(\rho(\xvec))
  + \edens_3(\rho(\xvec)) +\cdots \right\}
  \nonumber\\
  &=& 
  b_1\, {a_s^2\over 2M} \int \! \dthreex \, [\rho(\xvec)]^{7/3}
    \nonumber  \\ 
  & &
    \null +
    \left( b_2\, a_s^2 \, r_s +b_3\, a_p^3 +b_4\, a_s^3\right) 
     {1\over 2M} \int \! \dthreex \, [\rho(\xvec)]^{8/3}+\cdots
     \ ,
\eea
where the dimensionless $b_i$ are
\bea
b_1 &=& {4\over 35\pi^2}\, (g-1) \left({6\pi^2\over g}\right)^{4/3} 
    \, (11-2\ln2) \ ,\nonumber\\
b_2 &=& {1\over 10 \pi}\, (g-1)\left({6\pi^2\over g}\right)^{5/3}
    \nonumber \ ,\\ 
b_3 &=& {1\over 5\pi}\, (g+1)\left({6\pi^2\over g}\right)^{5/3}
    \nonumber \ ,\\
b_4 &=& \left({6\pi^2\over g}\right)^{5/3} 
     \biggl(  0.0755\,(g-1)+ 0.0574\,(g-1)(g-3)
\biggr)
     \ .  \label{eq:bfour}
\eea
In order to solve for the orbitals in
\Eq{eq:ks1} and to calculate the energy we need
the expression for $J_0(\xvec)$. 
In the LDA, this is simple since
\beq
J_0(\xvec) 
  =  \frac{\delta}{\delta\rho(\xvec)}
     \left( \til{\Gamma}_1[\rho] + \sum_{i=2}^\infty
             \til{\Gamma}_i[\rho] \right)
  = -{\delta\over\delta\rho(\xvec)}
     \left( E_{\rm HF}[\rho]+ E_{xc}[\rho]\right)
     \ .
\label{eq:j0dE}
\eeq
To NNLO we find:
\beq
J_0(\xvec) = -\frac{(g-1)}{g}\,\frac{4\pi\, a_s}{M}\,\rho(\xvec)
     - 
  {7\over3}\, b_1\, \frac{a_s^2}{2M}\, [\rho(\xvec)]^{4/3} -
  \frac{8}{3}\left( b_2\, a_s^2 \, r_s +b_3\, a_p^3 +b_4\, a_s^3\right) 
  \frac{1}{2M}[\rho(\xvec)]^{5/3}  \ .     
      \label{eq:J0solve}
\eeq
A convenient expression for the total binding energy (through NNLO)
follows by substituting for $J_0(\xvec)$ and combining terms: 
\bea
  E[\rho(\xvec)] &=& g \sum_i^{\rm occ.}\varepsilon_i
  - \frac{1}{2}\frac{(g-1)}{g}\,\frac{4\pi\, a_s}{M}
     \int\!\dthreex\,[\rho(\xvec)]^2
  - \frac{4}{3}\, b_1\, \frac{a_s^2}{2M}
     \int\!\dthreex\,[\rho(\xvec)]^{7/3}
     \nonumber \\
     & &
     \null - \frac{5}{3}
    \left( b_2\, a_s^2 \, r_s +b_3\, a_p^3 +b_4\, a_s^3\right) 
     \frac{1}{2M}  \int\!\dthreex\,[\rho(\xvec)]^{8/3}
     \ .
     \label{eq:Econ}
\eea
In the numerical calculations given below, the noninteracting case
($C_0=0$) uses $J_0(\xvec)\equiv 0$ and the first term in
Eq.~(\ref{eq:Econ}), LO uses the first term in Eq.~(\ref{eq:J0solve})
and the first two terms in Eq.~(\ref{eq:Econ}), and so on for NLO and
NNLO.

\subsection{Kohn-Sham Self-Consistent Procedure}

Here we describe the numerical procedure used to find the Kohn-Sham
orbitals.
We assume closed shells, so the density and potentials
are functions only of the radial coordinate $r \equiv |\xvec|$.
(This restriction is straightforward to relax.)
The Kohn-Sham iteration procedure is as follows:
 \begin{enumerate}
   \item 
   Guess an initial density profile $\rho(r)$.  If the system is
   particularly nonlinear this initial choice
   may be critical.  More generally, there may
   be a metastable state (e.g., if the system ultimately collapses). 
   For nuclear systems, the experience is that a crude caricature of the
   true density (such as a Woods-Saxon shape) 
   is adequate.  In the present case, the non-interacting
   harmonic oscillator density is sufficient.  

   \item 
   Evaluate the local single-particle potential 
   \beq
      v_s[\rho(r)] \equiv v_s(r) \equiv v(r) - J_0(r) 
        \label{eq:lspp}
   \eeq  
   using Eqs.~(\ref{eq:J0solve}) at the chosen level of
   approximation (e.g., NLO).
   Beyond the LDA $v_s$ could be a functional
   not just of the density but of the Kohn-Sham orbitals individually.

   \item 
   Solve the Schr\"odinger equation
   for the lowest $A$ states (including degeneracies), to find a set of
   orbitals and Kohn-Sham eigenvalues  
        $\{\varphi_\alpha,\varepsilon_\alpha\}$:
 \beq
  \bigl[ -\frac{{\nabla}^2}{2M}  +  v_s(r)
  \bigr]\, \varphi_\alpha({\bf x}) = \varepsilon_\alpha 
              \varphi_\alpha({\bf x})
  \, .
 \eeq

   \item 
   Compute a new density from the orbitals:
    \beq
      \rho(r) 
            = \sum_{\alpha=1}^{A} |\varphi_\alpha({\bf x})|^2 \ .
    \eeq
    All other ground-state observables are functionals of 
          $\{\varphi_\alpha,\varepsilon_\alpha\}$.

   \item
    Repeat steps 2.--4.\ until changes are acceptably
    small (``self-consistency'').  In practice, the changes in the
    density are ``damped'' by using a weighted average of
    the densities from the $(n-1)$th and $n$th iterations:
    \beq
       \rho(r) = \beta \rho_{n-1}(r) + (1-\beta) \rho_n(r)
       \ ,
    \eeq
    with $0 < \beta \le 1$.
  \end{enumerate}
  
This procedure has been implemented for dilute fermions in a trap using
computer codes written both in C and in Mathematica. 
Two methods for carrying out step 3.\ were tested. 
The Kohn-Sham single-particle equations are solved in one approach
by direct integration
of the differential equations via the Numerov method \cite{BLATT67} 
and in the other
approach 
by diagonalization 
of the single-particle Hamiltonian
in a truncated basis of unperturbed harmonic oscillator
wavefunctions.  The same results are obtained to high accuracy.
For closed shells, either method is efficient and easy to code.

\subsection{Fermions in a Harmonic Trap}

The interaction through NNLO is specified in terms of the three
effective range parameters $a_s$, $r_s$, and $a_p$. 
For the numerical calculations presented here, we consider two cases in the
dilute limit, $a_s \ll  \{r_s,a_p\} \approx 0$ with both signs for
$a_s$, and   
hard sphere repulsion with radius $R$, 
in which case $a_s=a_p=R$ and $r_s=2R/3$.

Lengths are measured in units of the oscillator
parameter $b \equiv \sqrt{\hbar/M\omega}$, 
masses in terms of the fermion mass $M$, and
$\hbar = 1$.
In these units, $\hbar\omega$ for the oscillator is unity and
the Fermi energy of a non-interacting gas with
filled shells up to $N_F$ is $E_F = (N_F + 3/2)$.
The total number of fermions $A$ is related to $N_F$ by
\beq
   A = \frac{g}{6}(N_F+1)(N_F+2)(N_F+3) \ .
\eeq
Since we have only considered spin-independent interactions, 
our results are independent of whether the spin degeneracy $g$ actually
originates from spin, isospin, or some flavor index.

With interactions included, single-particle states are labeled by a
radial quantum number $n$, an orbital angular momentum $l$ with
$z$-component $m_l$, and the spin projection.  The radial functions
depend only on $n$ and $l$, so the degeneracy of each level
is $g\times(2l+1)$.
Excluding spin, the solutions are of the form
\beq
   \varphi_{nlm_l} (\xvec) = \frac{u_{nl}(r)}{r}\, Y_{lm_l}(\Omega)
   \ ,
\eeq
where the radial function $u_{nl}(r)$ satisfies
\beq
 \left[
   -\frac{1}{2}\frac{d^2}{dr^2} + v_s(r) + \frac{l(l+1)}{2r^2} 
 \right] u_{nl}(r) = \varepsilon_{nl} u_{nl}(r)
  \ .
\eeq
The $u_{nl}$'s are normalized according to
\beq
   \int_0^\infty \! |u_{nl}(r)|^2\, dr = 1 \ .
\eeq
Thus the density is given by:
\beq
  \rho(r) = g\, \sum_i^{\rm occ.}\, 
     \frac{|u_{i}(r)|^2}{4\pi r^2}
     = g\, \sum_{nl}^{\rm occ.}\, 
       \frac{(2l+1)}{4\pi r^2}\,|u_{nl}(r)|^2 
\eeq
The interactions are sufficiently weak that the occupied states are in
one-to-one correspondence with those occupied in the non-interacting
harmonic oscillator potential.

\begin{table}
\renewcommand{\tabcolsep}{12pt}
\caption{\label{tab:TrapResults}%
Energies per particle, averages
of the local Fermi momentum $\kf$, and rms radii for a
variety of different parameters and particle numbers 
for a dilute Fermi gas in a harmonic trap.  
See the text for a description of units.
The effective range and p-wave scattering length are set
to zero, $r_s = a_p = 0$, except for the two lines where $a_s$ has
an asterisk, in which case they are given by $a_s = a_p = 3r_s/2$.}
\begin{ruledtabular}
\begin{tabular}{ccrddcdl}
  \multicolumn{1}{c}{$g$}  &  
  \multicolumn{1}{c}{$N_F$} &  
  \multicolumn{1}{c}{$A$} &  
  \multicolumn{1}{c}{$a_s$}   & 
  \multicolumn{1}{c}{$E/A$}  & 
  \multicolumn{1}{c}{$\langle \kf\rangle$} & 
  \multicolumn{1}{c}{$\sqrt{\langle r^2\rangle}$} & 
  approximation \\ \hline
  2 & 7 &  240 &  \thyp &  6.75  &  3.27  &  2.60  & $C_0 =0$ exact \\
  2 & 7 &  240 &  -0.16  &  5.98  &  3.61  &  2.35  & KS LO \\
  2 & 7 &  240 &  -0.16  &  6.25  &  3.44  &  2.47  & KS NLO (LDA) \\
  2 & 7 &  240 &  -0.16  &  6.23  &  3.46  &  2.46  & KS NNLO (LDA) \\
  \hline
  2 & 7 &  240 &   0.16  &  7.36  &  3.08  &  2.76  & KS LO \\
  2 & 7 &  240 &   0.16  &  7.51  &  3.03  &  2.81  & KS NLO (LDA) \\
  2 & 7 &  240 &   0.16  &  7.52  &  3.02  &  2.82  & KS NNLO (LDA) \\
  2 & 7 &  240 &   0.16^*  &  7.66  &  2.97  &  2.87  & KS NNLO (LDA) \\
  \hline
  4 & 4 &  140 &  \thyp &  4.50  &  2.66  &  2.12  & $C_0 =0$ exact \\
  4 & 4 &  140 &   -0.10  &  3.62  &  3.27  &  1.72  & KS LO \\
  4 & 4 &  140 &   -0.10  &  3.83  &  3.01  &  1.87  & KS NLO (LDA) \\
  4 & 4 &  140 &   -0.10  &  3.75  &  3.12  &  1.81  & KS NNLO (LDA) \\
  \hline
  4 & 4 &  140 &   0.10  &  5.09  &  2.44  &  2.31  & KS LO \\
  4 & 4 &  140 &   0.10  &  5.16  &  2.41  &  2.34  & KS NLO (LDA) \\
  4 & 4 &  140 &   0.10  &  5.18  &  2.40  &  2.35  & KS NNLO (LDA) \\
  4 & 4 &  140 &   0.10^*  &  5.20  &  2.39  &  2.36  & KS NNLO (LDA) \\
\end{tabular}
\end{ruledtabular}
\end{table}

\begin{figure}[p]
\centerline{\includegraphics*[width=9cm,angle=-90]{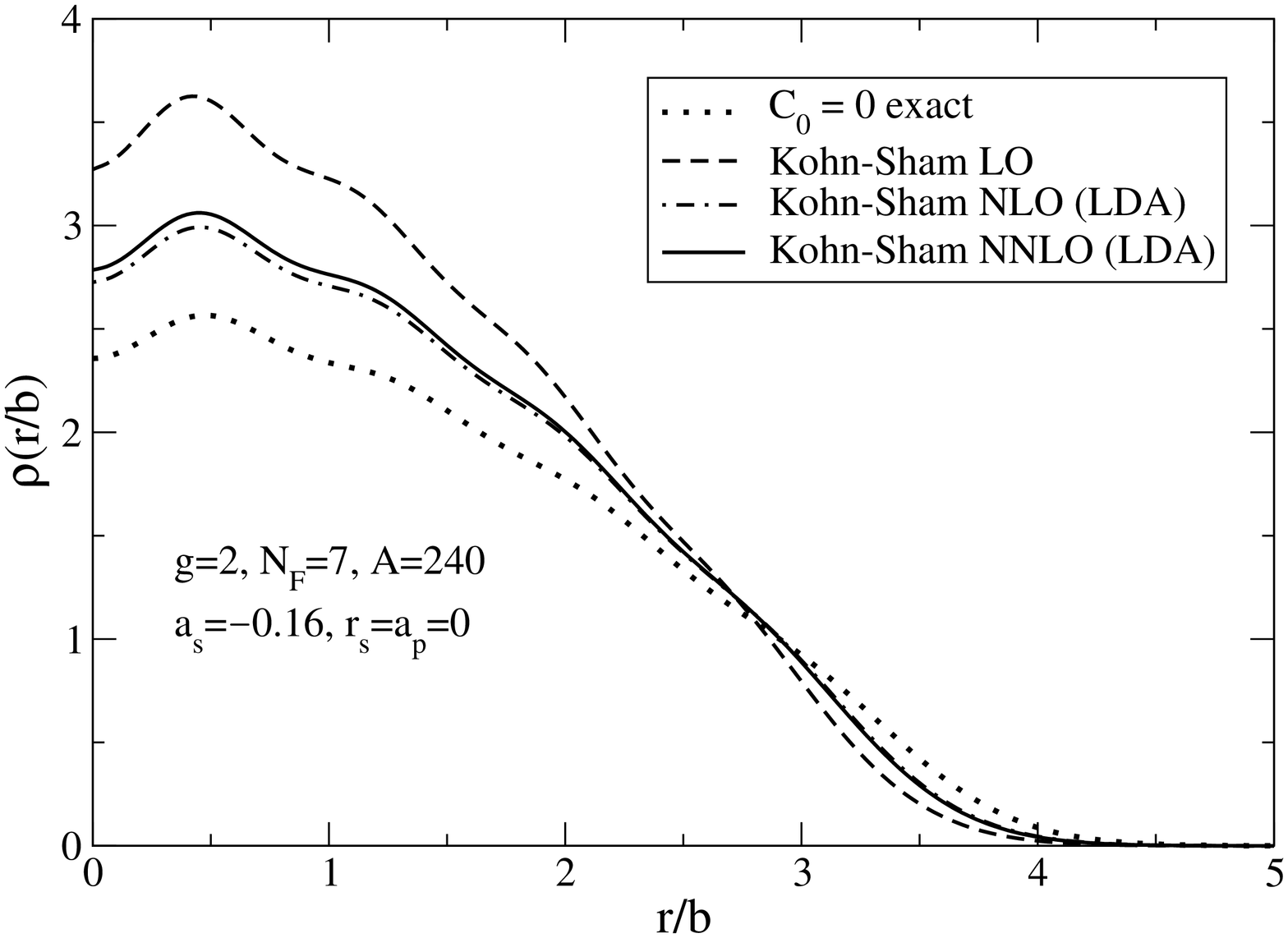}}
\vspace*{-.0in}
\caption{Kohn-Sham approximations (see text)
for a dilute gas of fermions in a
harmonic trap with degeneracy $g=2$ filled up to $N_F =7$, which implies
there are 240 particles in the trap.  
The scattering length is $a_s = -0.16$ and the other effective range
parameters are set to zero.}
\label{fig:ks1}
%
\vfill
%
\centerline{\includegraphics*[width=9cm,angle=-90]{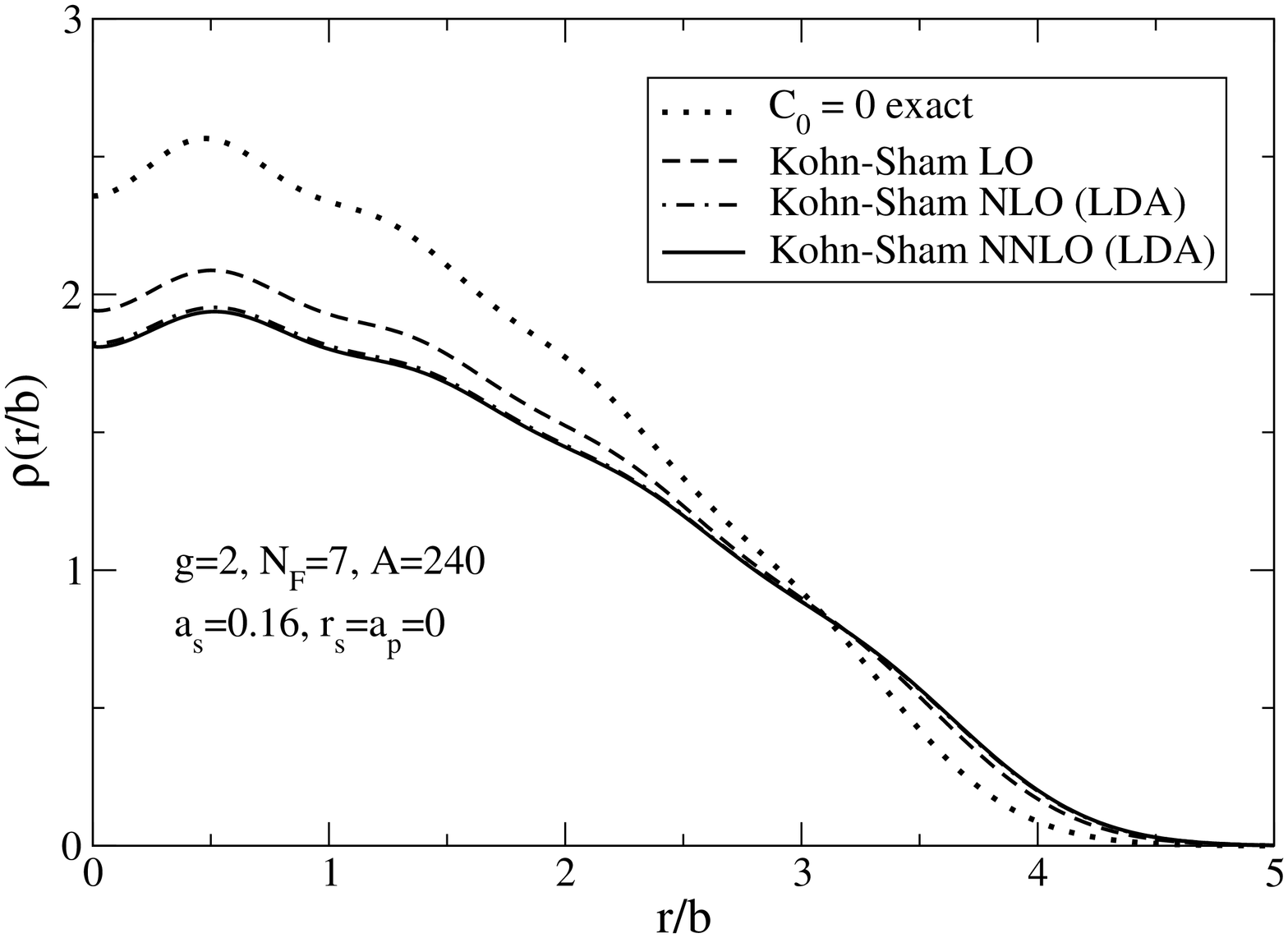}}
\vspace*{-.0in}
\caption{Kohn-Sham approximations (see text)
for a dilute gas of fermions in a
harmonic trap with degeneracy $g=2$ filled up to $N_F =7$, which implies
there are 240 particles in the trap.  
The scattering length is $a_s = +0.16$ and the other effective range
parameters are set to zero.}
\label{fig:ks2}
\end{figure}        
\begin{figure}[p]
\centerline{\includegraphics*[width=9cm,angle=-90]{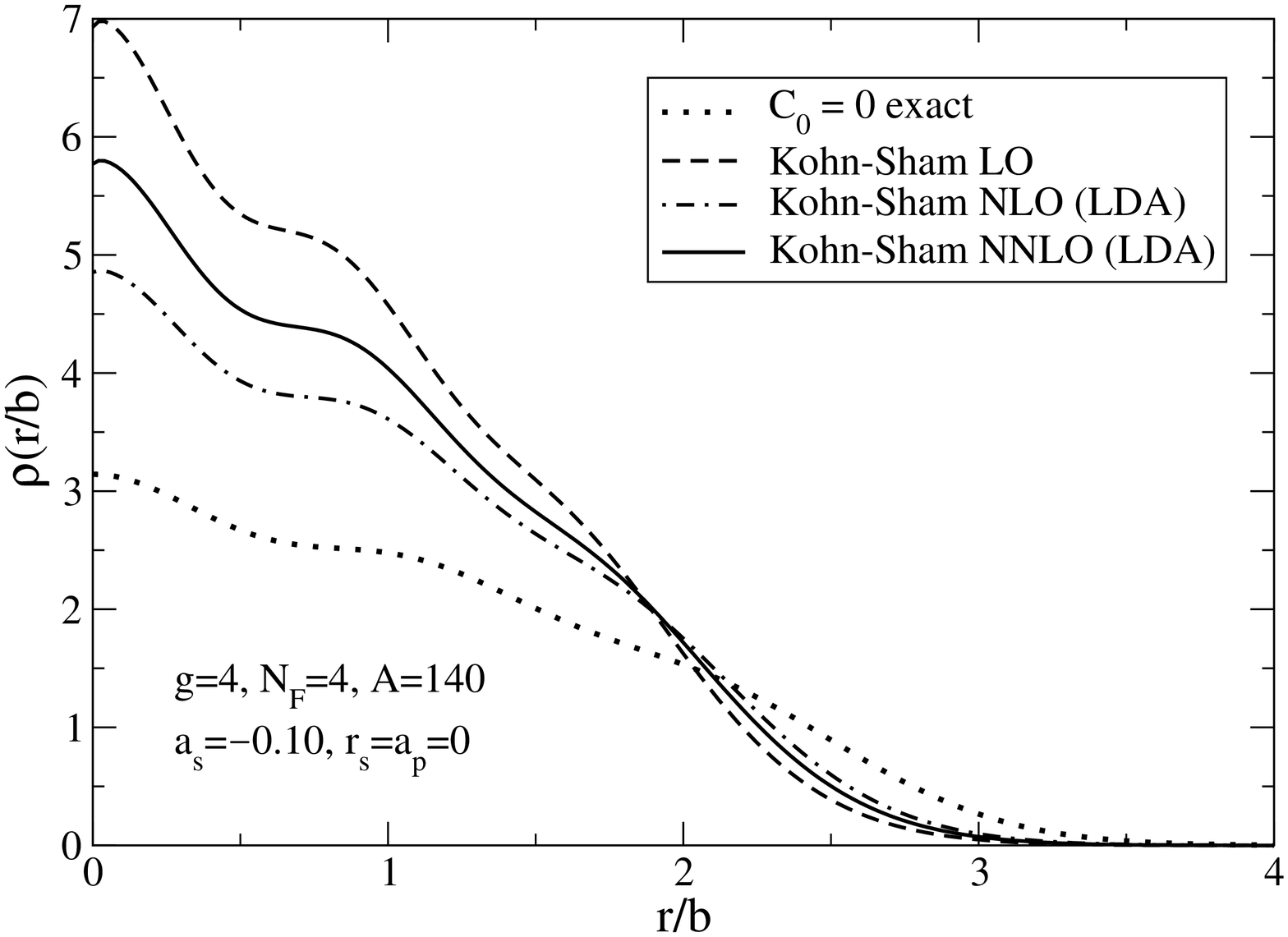}}
\vspace*{-.0in}
\caption{Kohn-Sham approximations (see text)
for a dilute gas of fermions in a
harmonic trap with degeneracy $g=4$ filled up to $N_F =4$, which implies
there are 140 particles in the trap.  
The scattering length is $a_s = -0.10$ and the other effective range
parameters are set to zero.}
\label{fig:ks5}
%
\vfill
%
\centerline{\includegraphics*[width=9cm,angle=-90]{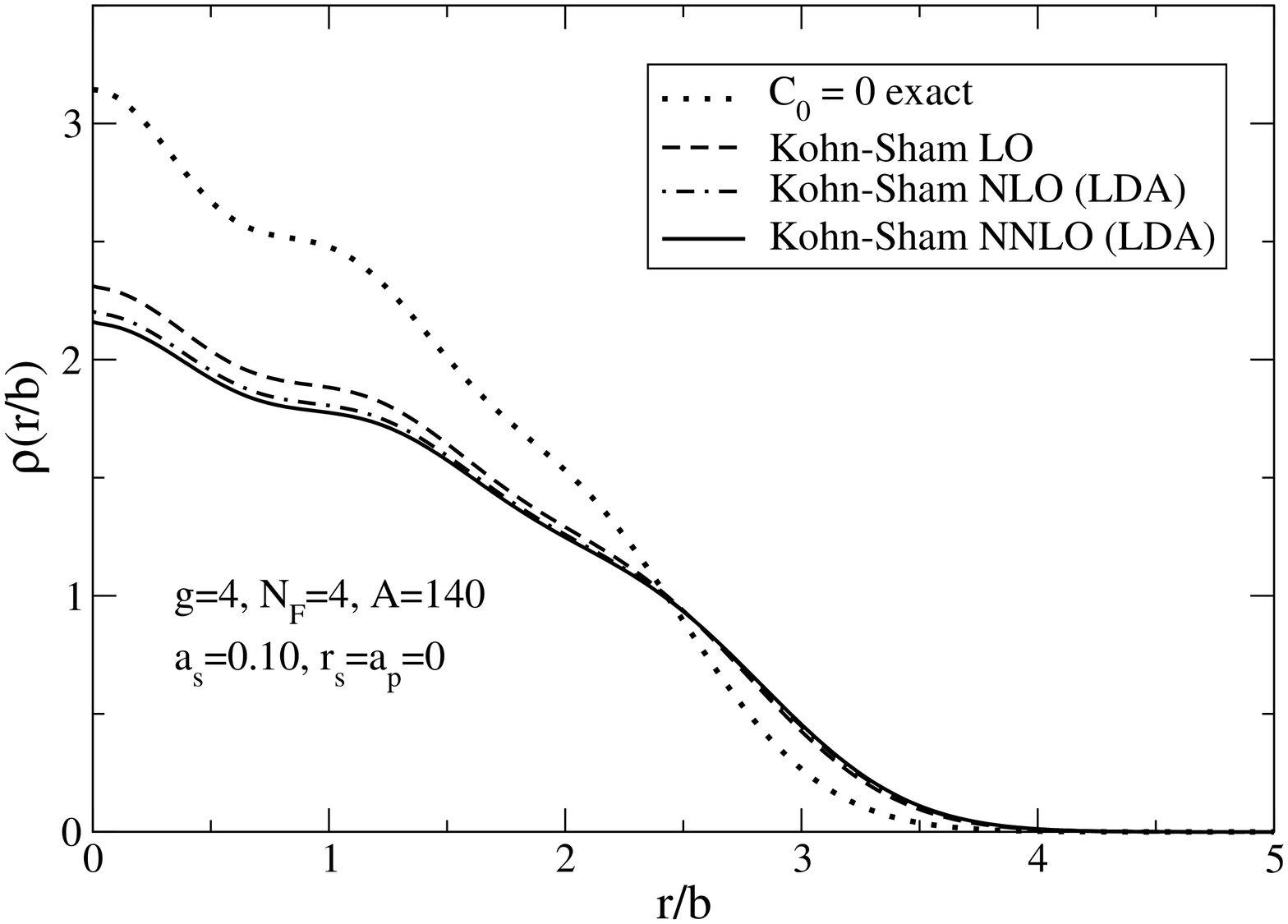}}
\vspace*{-.0in}
\caption{Kohn-Sham approximations (see text)
for a dilute gas of fermions in a
harmonic trap with degeneracy $g=4$ filled up to $N_F =4$, which implies
there are 140 particles in the trap.  
The scattering length is $a_s = +0.10$ and the other effective range
parameters are set to zero.}
\label{fig:ks6}
\end{figure}        
\begin{figure}[p]
\centerline{\includegraphics*[width=9cm,angle=-90]{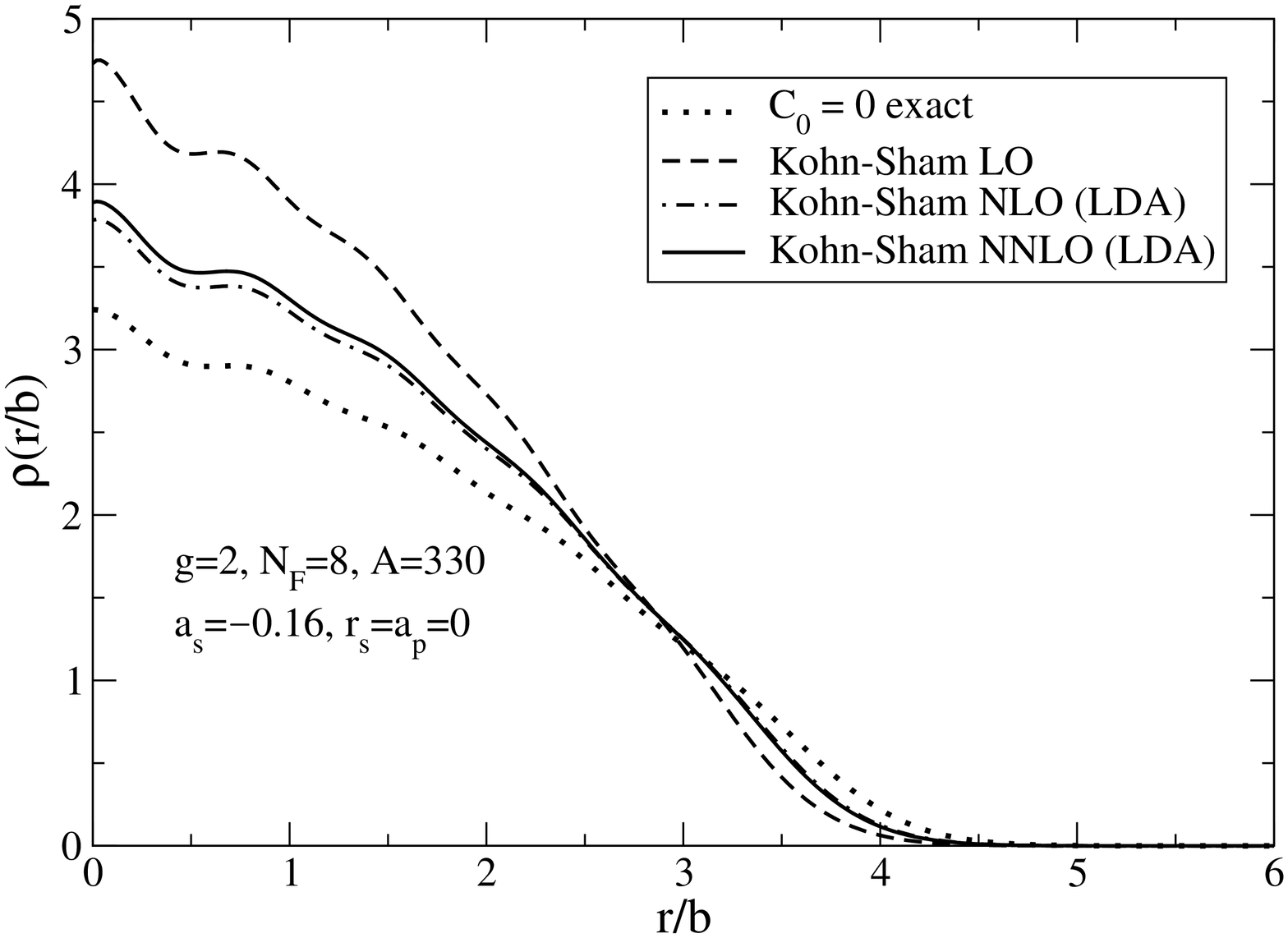}}
\vspace*{-.0in}
\caption{Kohn-Sham approximations (see text)
for a dilute gas of fermions in a
harmonic trap with degeneracy $g=2$ filled up to $N_F=8$, which implies
there are 330 particles in the trap.  
The scattering length is $a_s = -0.16$ and the other effective range
parameters are set to zero.}
\label{fig:ks3}
%
\vfill
%
\centerline{\includegraphics*[width=9cm,angle=-90]{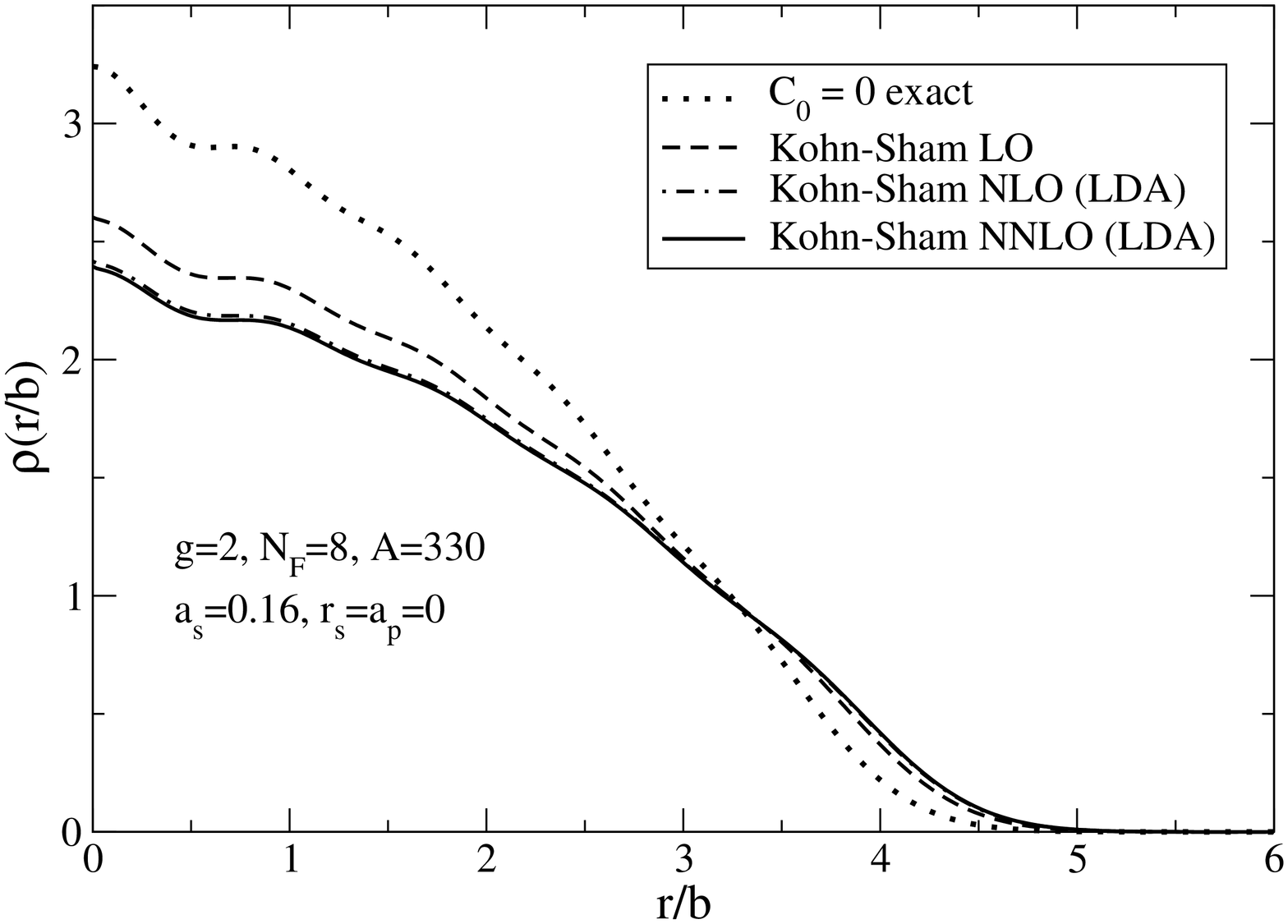}}
\vspace*{-.0in}
\caption{Kohn-Sham approximations (see text)
for a dilute gas of fermions in a
harmonic trap with degeneracy $g=2$ filled up to $N_F=8$, which implies
there are 330 particles in the trap.  
The scattering length is $a_s = +0.16$ and the other effective range
parameters are set to zero.}
\label{fig:ks4}
\end{figure}        
\begin{figure}[p]
\centerline{\includegraphics*[width=8.98cm,angle=-90]{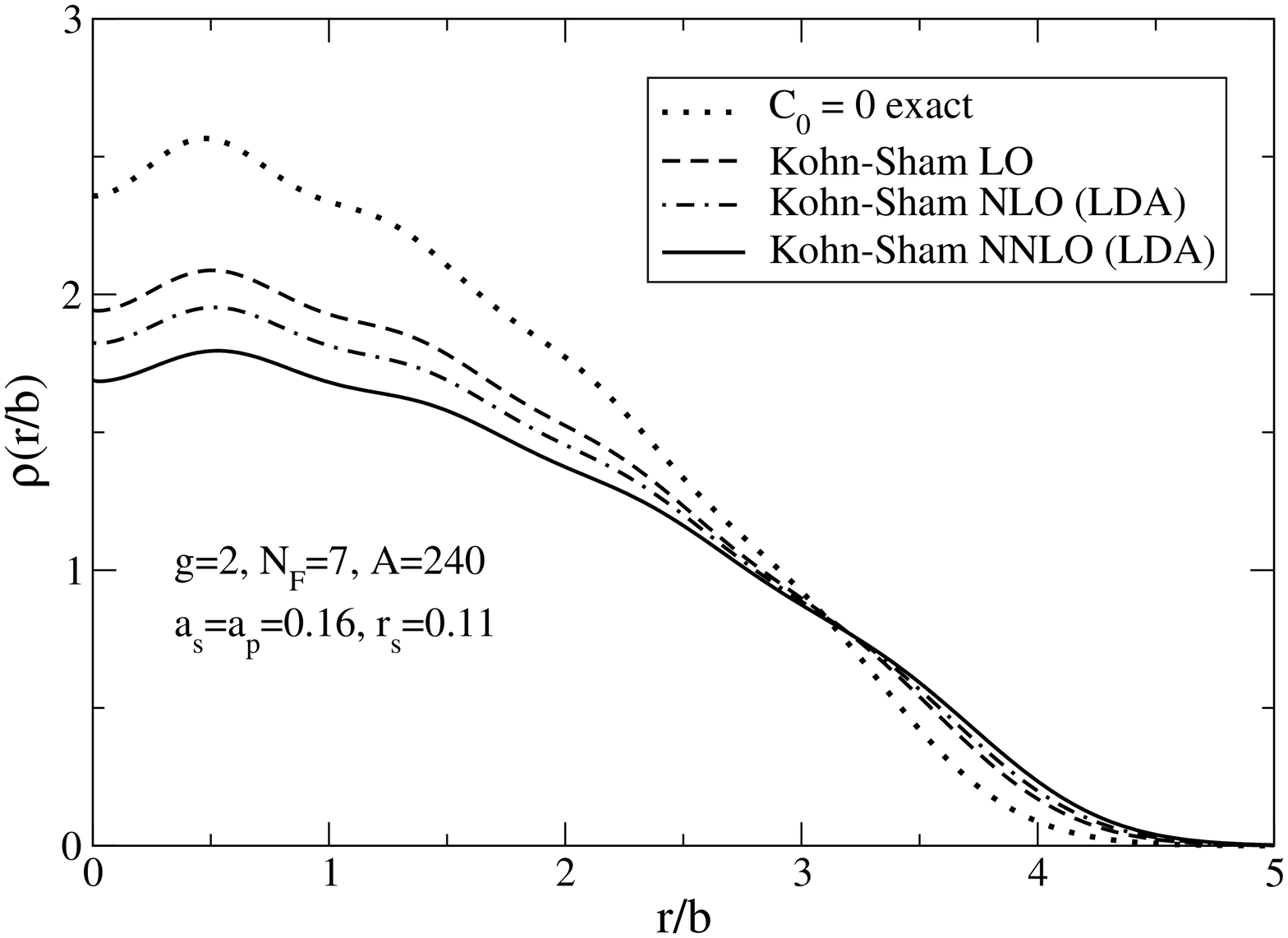}}
\vspace*{-.0in}
\caption{Kohn-Sham approximations (see text)
for a dilute gas of fermions in a
harmonic trap with degeneracy $g=2$ filled up to $N_F =7$, which implies
there are 240 particles in the trap.  
The scattering lengths are $a_s = a_p = 0.16$ and the effective range
is $r_s = 2a_s/3$ (hard sphere repulsion).}
\label{fig:ks5hs}
%
\vfill
%
\centerline{\includegraphics*[width=8.98cm,angle=-90]{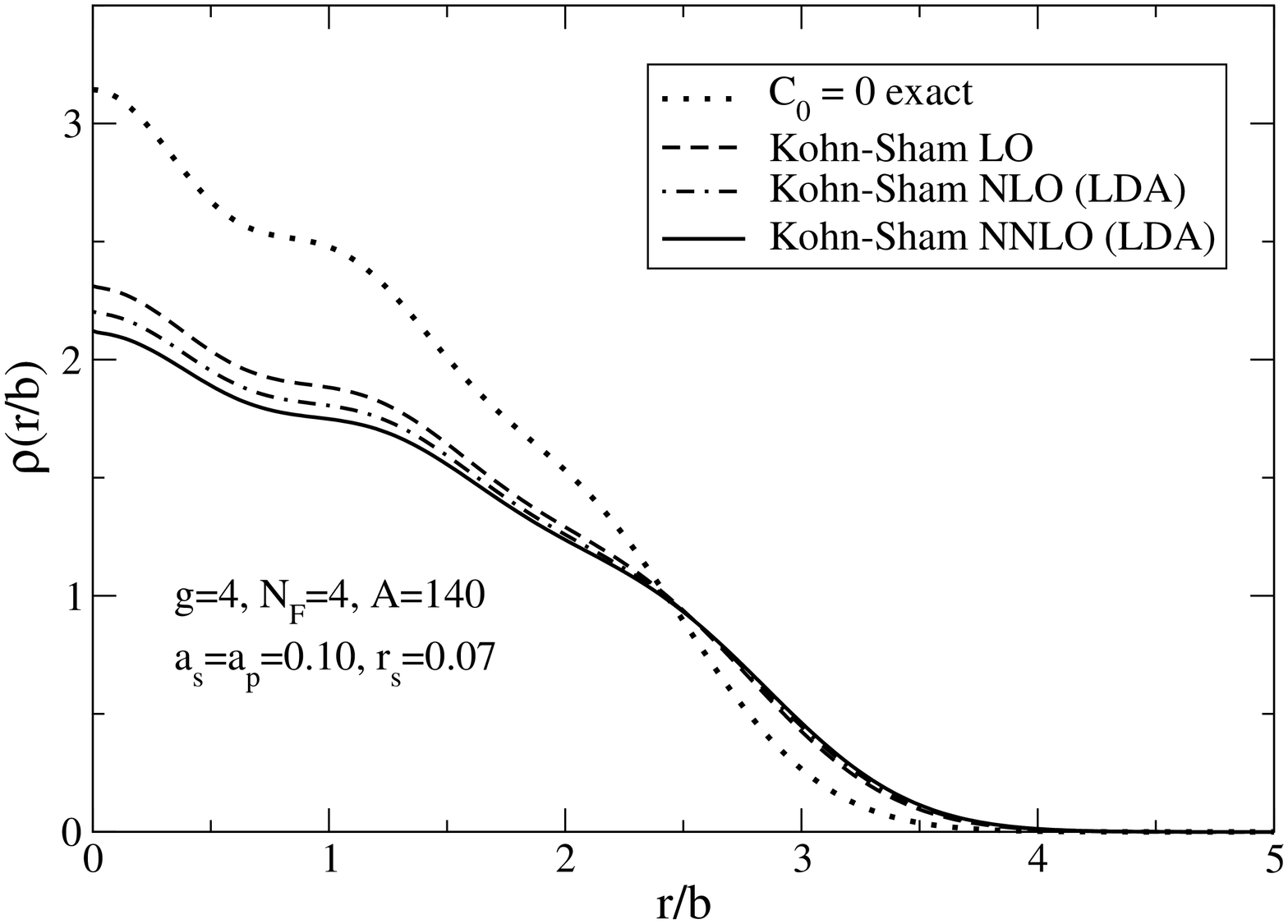}}
\vspace*{-.0in}
\caption{Kohn-Sham approximations (see text)
for a dilute gas of fermions in a
harmonic trap with degeneracy $g=4$ filled up to $N_F =4$, which implies
there are 140 particles in the trap.  
The scattering lengths are $a_s = a_p = 0.10$ and the effective range
is $r_s = 2a_s/3$ (hard sphere repulsion).}
\label{fig:ks6hs}
\end{figure}        

To check our numerical calculations, we first compared
to density distributions at zero temperature from Ref.~\cite{BRUUN98},
which used 
a contact interaction with strength corresponding to $a_s = -0.16$ in
our units.
In that work, non-interacting and mean-field (Hartree-Fock) results 
for the normal state (for possible comparison to superfluid solutions)
were presented for systems with 240 and 330 atoms.
The corresponding densities in our EFT expansion are the curves labeled
``$C_0 = 0$ exact'' and ``Kohn-Sham LO'' in Figs.~\ref{fig:ks1} and
\ref{fig:ks3}.
We also show densities for the same systems but with $a_s = +0.16$
in Figs.~\ref{fig:ks2} and \ref{fig:ks4}.
As one might expect, the attractive interaction pulls the density in
while the repulsive interaction pushes it out relative to the
non-interacting density.
Values for the energy per particle, the average Fermi momentum,
and the rms radius for each calculation are given
in Table~\ref{tab:TrapResults}.
Averages are defined as
\beq
   \langle f(\xvec) \rangle \equiv \frac{1}{A}
      \int\!\dthreex\, f(\xvec)\, \rho(\xvec) \ .
\eeq

In each of these figures, the other effective range parameters,
$r_s$ and $a_p$, are set to zero.
We have also shown results for a representative system with spin
degeneracy $g=4$ with weaker attractive ($a_s = -0.10$) and
repulsive ($a_s = +0.10$) interactions
in Figs.~\ref{fig:ks5} and \ref{fig:ks6}.
To illustrate the impact of $r_s$ and $a_p$ on the NNLO results,
we have also included two calculations where the underlying interaction
is hard-sphere repulsion with $R= 0.16$ ($g=2$) and $R=0.10$ ($g=4$)
in Figs.~\ref{fig:ks5hs} and \ref{fig:ks6hs}. 

From Figs.~\ref{fig:ks1} and \ref{fig:ks2}, we see what seems to be good
convergence of the density by NNLO.  Note that the LO results are not
well converged in either case.
From Table~\ref{tab:TrapResults}, we find that the average Fermi
momentum 
\beq
\langle\kf\rangle \equiv \frac{1}{A}
    \int\!\dthreex\, 
    \left[\frac{6\pi^2\rho(\xvec)}{g}\right]^{1/3}
    \, \rho(\xvec)
\eeq
is such that the expansion parameter
in an LDA sense, $\langle\kf\rangle a_s$, is equal to or greater than
one-half, which implies
poor convergence.
In fact, the convergence is misleading, as seen by comparison
to the hard-sphere repulsive case in Fig.~\ref{fig:ks5hs}.
(See the discussion of energy convergence in Sec.~\ref{sec:PCC}.)

The systems with $g=4$ have
$\langle\kf\rangle a_s \approx 1/4\mbox{--}1/3$, and
the convergence of even the hard-sphere case is good.
We note that the instability for large $g$ discussed in 
Ref.~\cite{Furnstahl:2002gt} is manifested as a collapse of the density
iteration by iteration when the stability criterion is violated by the
interior density.

\subsection{Comparison to Thomas-Fermi}
\label{subsec:TF}

In this section,
we compare the LDA Kohn-Sham (KS) results to those from a
Thomas-Fermi (TF) approximation.  By Thomas-Fermi, we mean
that the entire energy, including the kinetic energy, is calculated in a
local density approximation.   
In particular, the Thomas-Fermi kinetic energy functional
is \cite{ARGAMAN00}
\beq
  T_{TF}[\rho] = \frac{3}{10M}
    \left(\frac{6\pi^2}{g}\right)^{2/3}
    \int\!d^3r\, [\rho({\bf r})]^{5/3} 
    \ ,
\eeq
rather than being computed as the sum of the kinetic energy of Kohn-Sham
orbitals.
We define the Thomas-Fermi potential energy functional
to have the same form as
the Kohn-Sham potential energy functional in the LDA.
Historically, the Thomas-Fermi approach, as applied to atoms and
molecules, 
was the first attempt to use the (electron)
density as the basic variable rather than solving for the wavefunction.
In its original form, only the Hartree term and the external nuclear-electron
attractive potential were included in the potential energy.  We will
generalize to define LO, NLO, and NNLO Thomas-Fermi approximations.

The idea behind Thomas-Fermi is that in each volume element $dV$ we have
chemical equilibrium, with chemical potential $\mu$.
Each volume element is labeled by $r$ (we assume spherical symmetry for
convenience).
Local eigenvalues $E_k(r)$ are computed at each $r$
as if in a uniform system:
\beq
  E_k(r) = \frac{k^2}{2M} + v_s(r) \ ,
\eeq
and levels are filled until
\beq
  E_{\kf}(r) = \mu \ ,
\eeq
which defines the local Fermi momentum $\kf(r)$.
This is turn defines
the Thomas-Fermi density $\rho_{TF}(r,\mu)$ for a given chemical
potential $\mu$ as:
\beq
   \rho_{TF}(r,\mu) = \left\{
        \begin{array}{cl}
          \frac{\textstyle g}{\textstyle 6\pi^2}
             \bigl(2M[\mu - v_s(r)]  \bigr)^{3/2} 
          & \mbox{if $\mu > v_s(r)$} \\
          0  & \mbox{if $\mu \leq v_s(r)$}
        \end{array}
      \right.  \ ,
      \label{eq:rhoTF}
\eeq
where the potential $v_s(r)$ includes the external trap potential
and the LDA Kohn-Sham potential [see Eq.~(\ref{eq:lspp})].
The conventional TF procedure combines this equation with an equation
for the potential (e.g., a Poisson equation in the Coulomb case) by
substituting for the potential.  
Here we solve it in a two-step process closely analogous to the
Kohn-Sham solution procedure.

For a given choice of $\mu$, the number of fermions $A_{TF}$ is
given by:
\beq
   A_{TF}(\mu) = 4\pi \int_0^\infty\! r^2\, \rho_{TF}(r,\mu)\, dr\,
   ,    
\eeq  
where we've assumed a spherically symmetric distribution.
We find the correct value of $\mu$ for an input value of 
$A$ using a root finding
program, which finds the zero of
\beq
    f(\mu) \equiv A - A_{TF}(\mu)
\eeq
in the interval $\mu_{\rm min} < \mu < \mu_{\rm max}$.
The procedure to find the self-consistent $\rho_{TF}$ is iterative:
\begin{enumerate}
  \item Guess an initial $\rho_{TF}(r)$ (for example, the unperturbed
  density);
  \item given $\rho_{TF}(r)$, compute $v_s(r)$
    for the noninteracting, LO, NLO, or NNLO case;
  \item using this $v_s(r)$ in Eq.~(\ref{eq:rhoTF}), 
     find $\mu$ so that $f(\mu) = 0$;
  \item with the new value of $\mu$, calculate a new $\rho_{TF}(r,\mu)$
    from Eq.~(\ref{eq:rhoTF});
  \item return to step 2., continuing until $\rho_{TF}$ and $\mu$
    do not change within a prescribed tolerance.
  
\end{enumerate}

\begin{table}[t]
\renewcommand{\tabcolsep}{12pt}
\caption{\label{tab:TFResults}%
Comparisons of
energies per particle, averages
of the local Fermi momentum $\kf$, and rms radii
between Thomas-Fermi and Kohn-Sham treatments 
of a dilute Fermi gas in a harmonic trap.  
See the text for a description of units.
In all cases, $r_s = a_p = 0$.}
\begin{ruledtabular}
\begin{tabular}{ccrddcdl}
  \multicolumn{1}{c}{$g$}  &  
  \multicolumn{1}{c}{$N_F$} &  
  \multicolumn{1}{c}{$A$} &  
  \multicolumn{1}{c}{$a_s$}   & 
  \multicolumn{1}{c}{$E/A$}  & 
  \multicolumn{1}{c}{$\langle \kf\rangle$} & 
  \multicolumn{1}{c}{$\sqrt{\langle r^2\rangle}$} & 
  approximation \\ \hline
  2 & 2 &  20 &  \thyp &  3.00  &  2.15  &  1.73  & $C_0=0$ exact \\
  2 & 2 &  20 &  \thyp &  2.94  &  2.17  &  1.71  & $C_0=0$ TF \\
  2 & 2 &  20 &  -0.16  &  2.83  &  2.24  &  1.66  & KS NNLO (LDA) \\
  2 & 2 &  20 &  -0.16  &  2.77  &  2.26  &  1.64  & TF NNLO (LDA) \\
  2 & 2 &  20 &   0.16  &  3.22  &  2.04  &  1.83  & KS NNLO (LDA) \\
  2 & 2 &  20 &   0.16  &  3.15  &  2.06  &  1.81  & TF NNLO (LDA) \\
  \hline
  2 & 7 &  240 &  \thyp &  6.75  &  3.27  &  2.60  & $C_0=0$ exact \\
  2 & 7 &  240 &  \thyp &  6.72  &  3.29  &  2.59  & $C_0=0$ TF \\
  2 & 7 &  240 &  -0.16  &  6.23  &  3.46  &  2.46  & KS NNLO (LDA) \\
  2 & 7 &  240 &  -0.16  &  6.20  &  3.47  &  2.45  & TF NNLO (LDA) \\
  2 & 7 &  240 &   0.16  &  7.52  &  3.02  &  2.82  & KS NNLO (LDA) \\
  2 & 7 &  240 &   0.16  &  7.49  &  3.03  &  2.81  & TF NNLO (LDA) \\
  \hline
\end{tabular}
\end{ruledtabular}
\end{table}

This procedure is rather simple and scales very well with the number
of particles.
There are deficiencies to the Thomas-Fermi approach, however, for
many problems of interest.
These deficiencies  are most evident for the
Coulomb problem, where it is proved that molecules do not bind (the
separate atoms always have lower energy) \cite{TELLER62}.
Gradients expansions can improve the approximation but have not had a
quantitative impact compared to Kohn-Sham approaches.

The kinetic energy
contribution is a leading source of nonlocality in the energy
functional.   One of the chief virtues of the Kohn-Sham approach is that
it treats this component much more effectively than a derivative
expansion would.  The comparison with Thomas-Fermi calculations 
highlights this difference.
The most visible consequences are the absence of shell structure in
ground state densities in the Thomas-Fermi approach and
the incorrect treatment of the low-density asymptotic region (dominated
by the last Kohn-Sham orbitals).
In contrast, the last Kohn-Sham orbital has the correct energy (the
ionization energies in the exact and Kohn-Sham systems are equal), 
so the exponential tail of the distribution is correct.
These deficiencies of Thomas-Fermi are relevant for the calculation
of nuclear charge and matter densities.  

\begin{figure}[p]
\centerline{\includegraphics*[width=9cm,angle=-90]{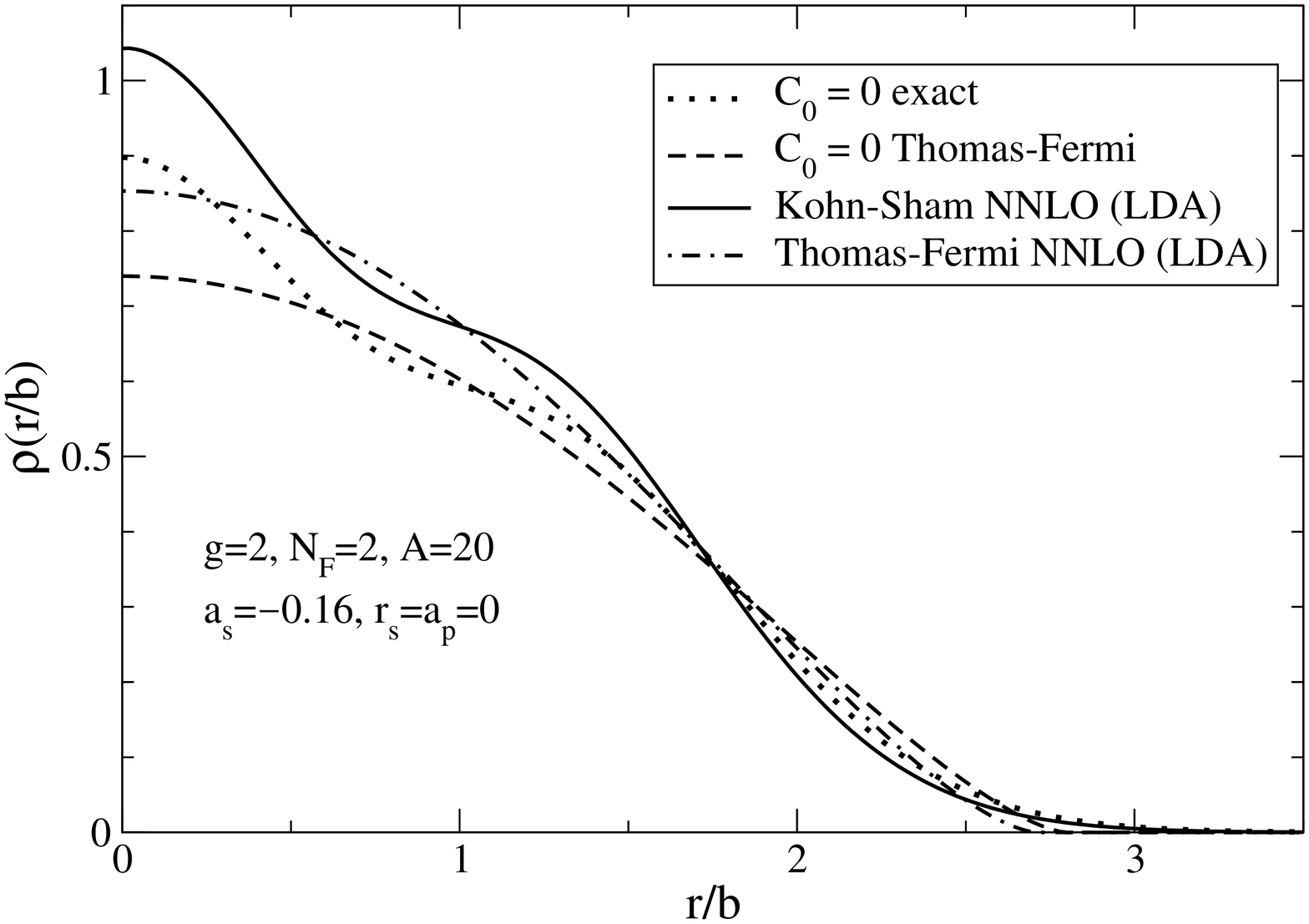}}
\vspace*{-.0in}
\caption{Thomas-Fermi and Kohn-Sham approximations for dilute gas of 
fermions in a
harmonic trap with degeneracy $g=2$ filled up to $N_F=2$, which implies
there are 20 particles in the trap.  
The scattering length is $a_s = -0.16$ and the other effective range
parameters are set to zero.}
\label{fig:tfks1}
%
%
\centerline{\includegraphics*[width=9cm,angle=-90]{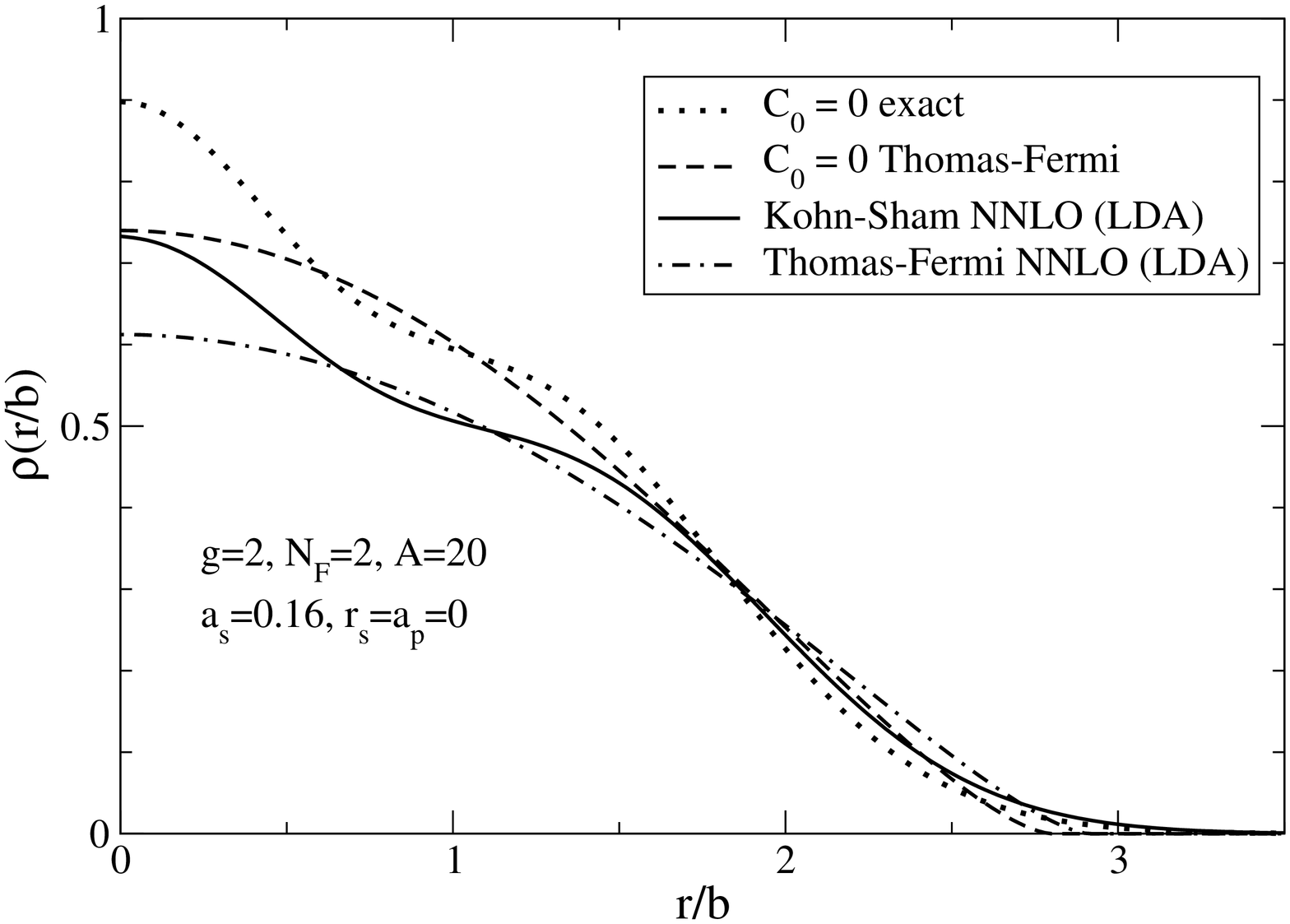}}
\vspace*{-.0in}
\caption{Thomas-Fermi and Kohn-Sham approximations for dilute gas of 
fermions in a
harmonic trap with degeneracy $g=2$ filled up to $N_F=2$, which implies
there are 20 particles in the trap.  
The scattering length is $a_s = +0.16$ and the other effective range
parameters are set to zero.}
\label{fig:tfks2}
\end{figure}        
\begin{figure}[p]
\centerline{\includegraphics*[width=9cm,angle=-90]{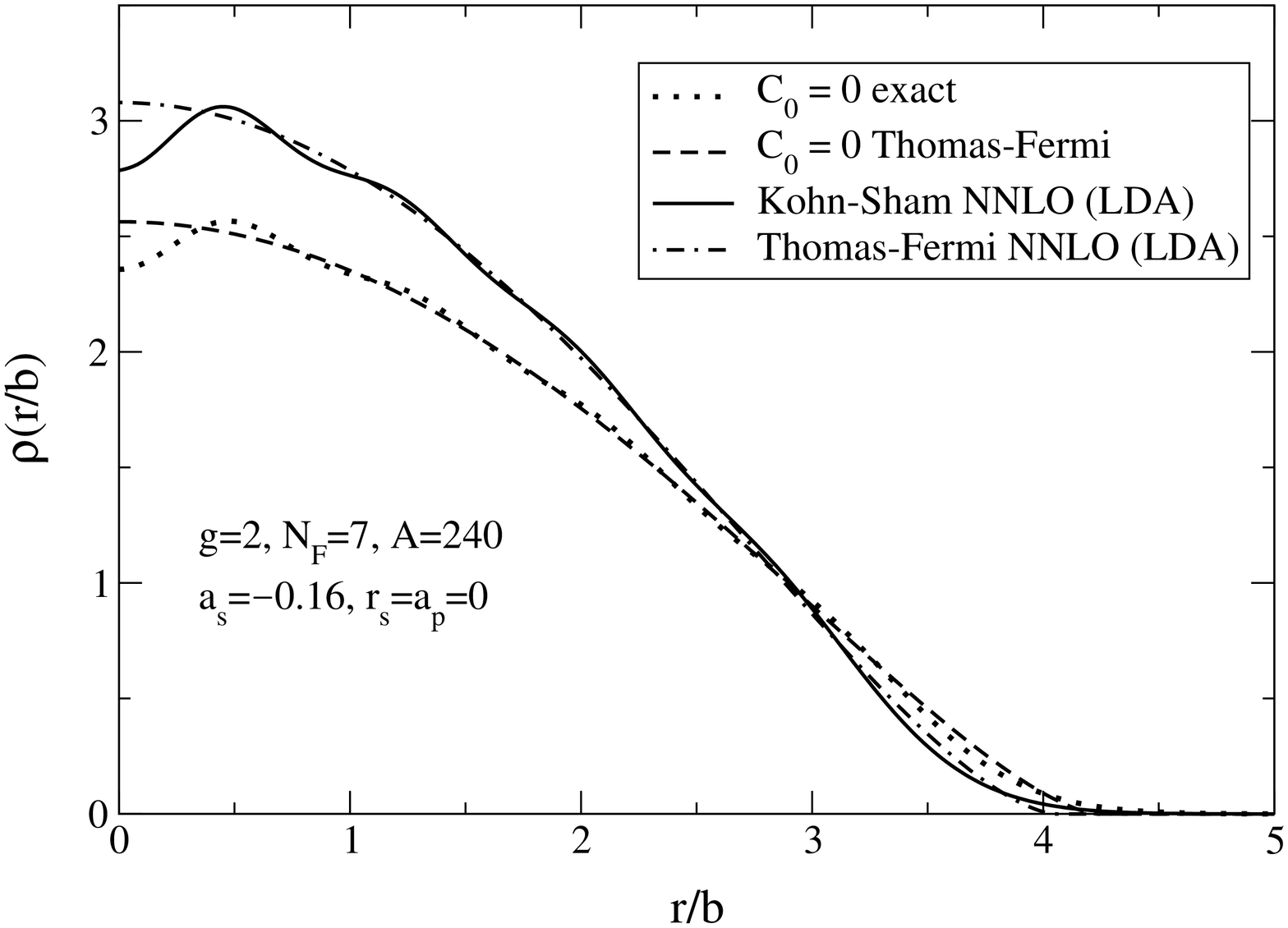}}
\vspace*{-.0in}
\caption{Thomas-Fermi and Kohn-Sham approximations for dilute gas of 
fermions in a
harmonic trap with degeneracy $g=2$ filled up to $N_F=7$, which implies
there are 240 particles in the trap.  
The scattering length is $a_s = -0.16$ and the other effective range
parameters are set to zero.}
\label{fig:tfks3}
\vfill
\centerline{\includegraphics*[width=9cm,angle=-90]{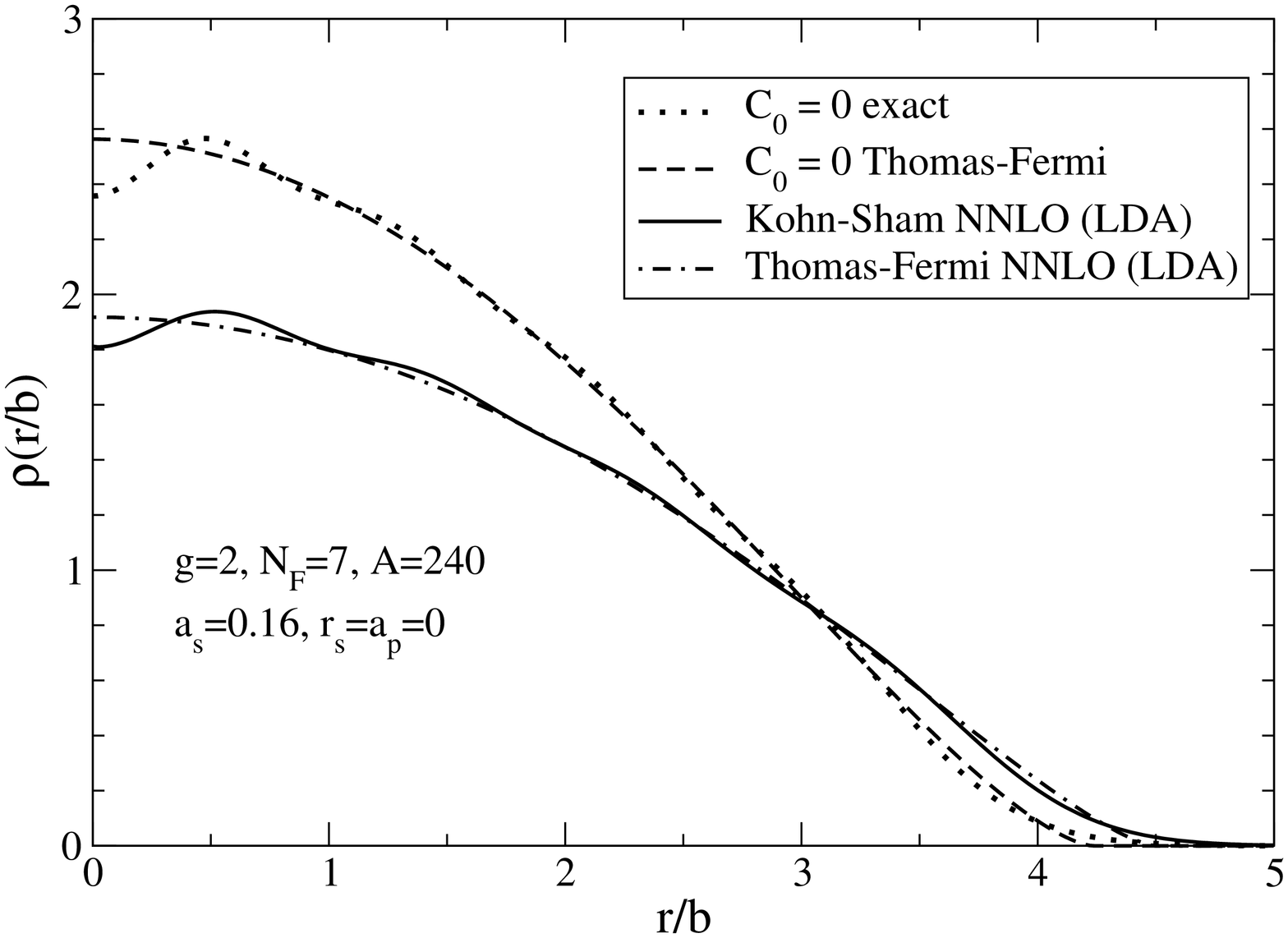}}
\vspace*{-.0in}
\caption{Thomas-Fermi and Kohn-Sham approximations for dilute gas of 
fermions in a
harmonic trap with degeneracy $g=2$ filled up to $N_F=7$, which implies
there are 240 particles in the trap.  
The scattering length is $a_s = +0.16$ and the other effective range
parameters are set to zero.}
\label{fig:tfks4}
\end{figure}        

In Figs.~\ref{fig:tfks1} and \ref{fig:tfks2}, 
Thomas-Fermi and Kohn-Sham densities are
compared for a small system of 20 atoms (the other parameters are given
in the figures).  Both non-interacting and NNLO curves are shown in each
figure.  
The shell structure is evident in the non-interacting density and
is only slightly damped by the interactions in the Kohn-Sham
approach.  
In contrast, the Thomas-Fermi curves are featureless.
The same comparisons but with an order of magnitude more atoms ($A=240$) are
shown in Figs.~\ref{fig:tfks3} and \ref{fig:tfks4}.
In these case the differences are much smaller.
For a small number of particles, the Kohn-Sham procedure is a comparable
computation to Thomas-Fermi.
With thousands of atoms, the Thomas-Fermi approximation should be
accurate and efficient.

In Table~\ref{tab:TFResults}, the energies per particle 
and other properties are compared for Kohn-Sham and Thomas-Fermi
solutions.  
With $A=20$ atoms, the energy is reproduced at about the 2\% level
and the average Fermi momentum at the 1\% level.
With $A=240$ atoms, the energy is reproduced better than 1\%
and the average Fermi momentum to a third of a percent.
Thus the Thomas-Fermi density will be quite adequate in making 
error estimates, which we turn to next.

\subsection{Power Counting and Convergence}
\label{sec:PCC}

A major motivation for the use of effective field theory for
many-body problems is the promise of error estimates.
We face two challenges in making such estimates for a Kohn-Sham DFT. 
There are usually new low-energy constants (LEC's) at each
successive order in the EFT expansion.  
We first need to estimate the size of the LEC's in the omitted
orders.
Second, we need to estimate their numerical impact on the functional and
subsequently on observables in finite systems.
In the present case, where we have a perturbative low-density
expansion, we can
do both directly.

Naive dimensional analysis, or NDA, is frequently used to estimate
unknown
LEC's in an EFT Lagrangian.
The idea is to identify the relevant dimensional momentum and mass
scales in the problem and to rescale each term in the Lagrangian as an
appropriate combination of scales times a dimensionless coefficient. 
An estimate of the truncation error follows by assuming the
dimensionless coefficient is order unity (e.g., from 1/3 to 3).

In Ref.~\cite{FURNSTAHL99b}, NDA appropriate to low-energy chiral
effective field theories of QCD was applied to covariant 
energy functionals for
nuclei.  While these functionals were viewed as approximate Kohn-Sham
energy functionals, their form was directly derived from a chiral
Lagrangian in the Hartree approximation.  
That is, there was a one-to-one correspondence between terms in
the Lagrangian and terms in the corresponding energy functional.
As a result, estimates of
coefficients in the Lagrangian were immediately translated into error
estimates in the functional, which led to estimates in specific finite
nuclei through local density approximations \cite{FURNSTAHL99b}. 

In the present case, the Hartree (actually Hartree-Fock) 
terms again provide an immediate
connection between NDA estimates in the Lagrangian and estimates of 
the energy per particle in terms of the fermion density.  Using LDA
estimates of the average density, we get energy estimates for specific
systems of trapped atoms.
EFT power counting then provides estimates for the higher-order terms
using the
normalization established by the Hartree terms.
For a purely short-ranged interaction, the NDA is simple: the estimate
of the Hartree-Fock
energy contribution from a given term in the Lagrangian
is found by replacing $\psi^\dagger\psi$ by the average density
(and including an appropriate spin factor).  Any reasonable density
(e.g., the Thomas-Fermi result) can be used to find the average density,
since we already allow for a much larger uncertainty in the coefficients.
Local density estimates for derivatives of densities are also
sufficiently accurate.

\begin{figure}[t]
\centerline{\includegraphics*[width=9cm,angle=0]{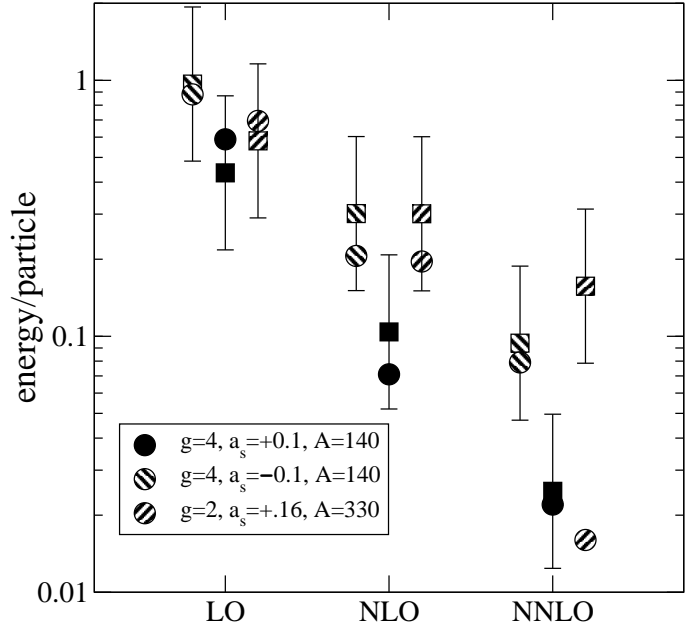}}
\vspace*{-.0in}
\caption{Contributions to the energy per particle for three example
systems of atoms in a harmonic trap.  The actual contribution in each
case is shown as a round symbol.  The square symbols are estimates based
on naive dimensional analysis (see text), with error bars marking 
a range from 1/2 to 2 in the dimensionless coefficients.}
\label{fig:error_plot}
\end{figure}        

In Fig.~\ref{fig:error_plot}, the contributions to the energy per
particle from LO, NLO, and NNLO terms in the energy per particle
are shown as round symbols for three of the 
systems from Table~\ref{tab:TrapResults}.
So, for example, the NLO contribution is from 
$|E_{\rm NLO}-E_{\rm LO}|/A$.
The square symbol denote estimates based on naive dimensional analysis,
with error bars indicating a 1/2 to 2 uncertainty in the coefficients.
In particular, the LO contribution is an estimate of the Hartree-Fock 
term and the NLO and NNLO estimates were found by multiplying the LO
result by $\langle\kf\rangle a_s$ and $(\langle\kf\rangle a_s)^2$,
respectively.
With one exception, the NDA estimates are good predictors of the
actual contribution.  Therefore, we can use these estimates to predict
the uncertainty in the energy per particle from higher orders.
The exception is the NNLO estimate for the $g=2$ system, which
greatly overestimates the actual contribution at that order.
The reason is evident from the $b_4$ coefficient in Eq.~(\ref{eq:bfour}), 
which determines the
size of this contribution.   Each of the two terms in $b_4$
are the size expected from NDA, but for $g=2$ they happen to largely cancel.
There is always this possibility of unnaturally small coefficients
(and subsequent contributions) because of accidental cancellations.
If we compare instead the NNLO energy from the hard-sphere-repulsion
case, which has additional contributions at NNLO, we find that it is
close to the NDA estimate.


\section{Summary and Conclusions}
\label{sect:summary}

In this paper,
we construct a Kohn-Sham density functional for a confined, dilute Fermi
gas as an order-by-order effective field theory (EFT) expansion.
The starting point is 
a generating functional with a source $J(x)$ coupled
to the composite density operator $\psi^\dagger\psi$.   
A functional Legendre transformation with respect to the source leads to
an effective action of the density.
In general we might expect complications
with such an effective action because of the renormalization
of composite operators \cite{YOKOJIMA95,OKUMURA96}, 
but these are avoided in the present case because the
fermion number is a conserved charge \cite{COLLINS86}.

A conventional effective action is a functional 
of the expectation value of elementary fields in
the Lagrangian.
It defines a classical field theory that contains all of the
quantum effects of the interacting quantum field theory associated with
the original Lagrangian.  That is, 
one reproduces results of the full field theory from tree
level calculations based on the propagators and vertices of the
effective action \cite{COLEMAN88,PESKIN95,WEINBERG96}.  
In the present case, the effective action of the
density can be used to calculate the ground state energy, including all
correlations, with what looks like a Hartree calculation (which is the
many-body equivalent of tree level).  This is density functional theory.

The calculation is carried out by adapting the inversion method proposed
in Refs.~\cite{FUKUDA94,FUKUDA95} and the Kohn-Sham
procedure of Ref.~\cite{VALIEV97} to an EFT treatment of the dilute
Fermi gas.  Instead of organizing the perturbative inversion in terms of
a coupling constant (e.g., the electron charge squared), we use the EFT
expansion parameter for the natural dilute system, which is $1/\Lambda$,
where $\Lambda$ is the resolution scale.  In the uniform system, the
ultimate dimensionless expansion parameters are products of the Fermi
momentum $\kf$ and parameters of the effective range expansion (e.g.,
the scattering length $a_s$, which is of order $1/\Lambda$ for
a natural system).  
In a finite system, the density-weighted average of a local Fermi
momentum times the effective range parameters controls one expansion,
with another expansion involving gradients of the density.

In the present work, we used the local density approximation (LDA),
which means that only the first expansion was tested.
The observed
convergence of the density and energy in sample systems 
confirms this expansion for a finite system.
An error plot of contributions to the energy per particle versus the
order of the calculation shows that we can reliably estimate the
truncation error in a finite system.
The next step is to develop a derivative expansion and test it for
convergence. 
Work is in progress to adapt to the present problem
the methods that have been used to derive derivative
expansions for one-loop effective actions.

The EFT expansion for the uniform system using dimensional
regularization and minimal subtraction (DR/MS) offered many simplifications
over conventional treatments \cite{HAMMER00}.  
The derivative expansion approach would preserve all of these
advantages.
In particular, renormalization of the derivative expansion coefficient
functions would be carried out entirely in the
uniform system using DR/MS, so that a given diagram contributes to
only a single prescribed order in the expansion. 

An alternative to the inversion method that might be preferred for some
systems is to introduce an auxiliary field (or fields) and to carry out
the Legendre transformation conventionally with respect to 
that field \cite{FUKUDA94,FUKUDA95}.
This could be applied to the large-N EFT expansion discussed in
Ref.~\cite{Furnstahl:2002gt}.  The construction of  Kohn-Sham DFT in the
auxiliary field formulation is discussed in
Refs.~\cite{FAUSSURIER00,VALIEV96}.

Many of the phenomena of greatest interest in experiments on trapped
fermion atoms involve tuning the system so that the $s$-wave 
scattering length is
large.  For example, one would like to study superfluid 
transitions at low temperature \cite{GRANADE02,HOLLAND01,TIMMERMANS01}.
systematic solution to 
the large scattering length problem at finite density is not yet
available, even for a uniform system.
Furthermore, with \emph{any} attractive interaction we expect a pairing
instability.  
Extending the DFT/EFT expansion procedure to include pairing will be
an important challenge.
The inversion method has been applied to conventional BCS
superconductivity in Ref.~\cite{INAGAKI92} using a source
coupled to the pair creation and destruction operator.
Work to adapt this procedure to the EFT is in progress.

These extensions will be directly relevant for nuclear applications as
well.
In addition,
to adapt the density functional procedure to chiral effective field
theories with explicit pions, 
we will need to extend the discussion to include long-range
forces (``long range'' means compared to $1/\Lambda$).
Other extensions include DFT with  alternative source terms
(e.g., spin-density DFT) and time-dependent DFT in the EFT formalism.
Studies of each of these extensions are in progress.

\acknowledgments

We thank P.~Bedaque, 
H.-W.\ Hammer, A.~Schwenk and B.D.\ Serot  for useful comments.
This work was supported in part by the National Science Foundation
under Grant No.~PHY--0098645.



\begin{thebibliography}{99} 
\bibitem{KOHN65} W. Kohn and L. J. Sham, Phys.\ Rev.\  {\bf A140} (1965) 1133.

\bibitem{PARR89}R.~G.\ Parr and W. Yang, {\it Density Functional Theory of Atoms
and Molecules} (Oxford University Press, New York, 1989)

\bibitem{DREIZLER90}R.~M.\ Dreizler and E.~K.~U.\ Gross,
      {\it Density Functional Theory} (Springer, Berlin, 1990).

\bibitem{KOHN99}W. Kohn, Rev.\ Mod.\ Phys.\ {\bf 71} (1999) 1253.
      
\bibitem{ARGAMAN00}N. Argaman and G. Makov, Amer.\ J.\ Phys.\ {\bf 68} 69.         

\bibitem{PETKOV91} I.~Zh.\ Petkov and M.~V.\ Stoitsov, {\it Nuclear Density Functional Theory}
(Clarendon Press, Oxford, 1991)

\bibitem{EFT98} 
Proceedings of the Joint Caltech/INT Workshop: {\em Nuclear
Physics with Effective Field Theory}, ed.\ R.~Seki, U.~van Kolck, and
M.~J.~Savage (World Scientific, 1998).

\bibitem{EFT99} 
Proceedings of the INT Workshop: {\em Nuclear
Physics with Effective Field Theory II}, ed.\ P.F.~Bedaque, M.J.~Savage,
 R.~Seki, and U.~van Kolck
 (World Scientific, 2000).

\bibitem{BEANE99} S.R.~Beane, P.F.~Bedaque, W.C.~Haxton, D.R.~Phillips,
                  and M.J.~Savage,\\
                 ``From Hadrons to Nuclei: Crossing the Border'',
                  {\tt [nucl-th/0008064]}.

\bibitem{Birareview} U. van Kolck, Prog. Part. Nucl. Phys. {\bf 43} 
(1999) 337.

\bibitem{Furnstahl:2001hs}
R.~J.~Furnstahl,
arXiv:nucl-th/0109007.

  
%
\bibitem{PERDEW96}J.~P.\ Perdew, K.~Burke, and M.~Ernzerhof,
  Phys.\ Rev.\ Lett.\ {\bf 77} (1996) 3865; {\bf 78} (1997) 1396(E).

\bibitem{PERDEW99}J.~P.\ Perdew, S.~Kurth, A.~Zupan, and P.~Blaha,
  Phys.\ Rev.\ Lett.\ {\bf 82} (1999) 2544.



\bibitem{COLEMAN88}S.~Coleman, {\it Aspects of Symmetry\/}
   (Cambridge Univ.\ Press, New York, 1988).

\bibitem{WEINBERG96} S.~Weinberg, {\it The Quantum Theory of
     Fields:~vol.~II, Modern Applications} (Cambridge University 
     Press, 1996).

\bibitem{PESKIN95}M.E.~Peskin and D.V.~Schroeder,
 {\it An Introduction to Quantum Field Theory} (Addison--Wesley, 1995).


\bibitem{GRANADE02}S.~R.\ Granade, M.~E.\ Gehm, K.~M.\ O'Hara,
  and J.~E.\ Thomas, Phys.\ Rev.\ Lett.\ {\bf 88} (2002) 120405.
      
\bibitem{KOCH00}W.~Koch and M.~C.\ Holthausen, {\it A Chemist's Guide
   to Density Functional Theory} (Wiley-VCH, New York, 2000.

\bibitem{BRACK85}M.~Brack, Helv.\ Phys.\ Acta {\bf 58} (1985) 715.

\bibitem{SCHMID95}R.~N.\ Schmid, E.~Engel, and R.~M.\ Dreizler,
  Phys.\ Rev.\ C {\bf 52} (1995) 164.

\bibitem{SCHMID95a}R.~N.\ Schmid, E.~Engel, and R.~M.\ Dreizler,
  Phys.\ Rev.\ C {\bf 52} (1995) 2804.

\bibitem{FUKUDA94}R.~Fukuda, T.~Kotani, Y.~Suzuki, and S.~Yokojima,
       Prog.\ Theor.\ Phys.\ {\bf 92} (1994) 833.
       
\bibitem{FUKUDA95}R.~Fukuda, M.~Komachiya, S.~Yokojima, Y.~Suzuki,
  K.~Okumura, and T.~Inagaki, Prog.\ Theor.\ Phys.\ Suppl.\
    {\bf 121} (1995) 1.


\bibitem{VALIEV97}M.\ Valiev and G.~W.\ Fernando,
    arXiv:cond-mat/9702247 (1997), unpublished.
    
\bibitem{VALIEV97b}M.\ Valiev and G.~W.\ Fernando,
    Phys.\ Lett.\ A {\bf 227} (1997) 265.

\bibitem{RASAMNY98}M.~Rasamny, M.~M.\ Valiev, and G.~W.\ Fernando,
    Phys.\ Rev.\ B {\bf 58} (1998) 9700.
    
\bibitem{VALIEV96}M.\ Valiev and G.~W.\ Fernando,
   Phys.\ Rev.\ B {\bf 54} (1996) 7765.
   
\bibitem{FAUSSURIER00}G.~Faussurier, J.\ Quant.\ Spect.\ Rad.\ Trans.\
  {\bf 65} (2000) 207.
   
\bibitem{CHITRA00}R.~Chitra and G.~Kotliar, Phys.\ Rev.\ B
   {\bf 62} (2000) 12715.
   
\bibitem{CHITRA01}R.~Chitra and G.~Kotliar, Phys.\ Rev.\ B
   {\bf 63} (2001) 115110.

\bibitem{Polonyi:2001uc}
J.~Polonyi and K.~Sailer,
arXiv:cond-mat/0108179.


\bibitem{LEPAGE89} G. P. Lepage,
  ``What is Renormalization?'', 
        in {\it From Actions to Answers} (TASI-89), edited by
        T. DeGrand and D. Toussaint (World Scientific, Singapore, 1989);
         ``How to Renormalize the Schr\"odinger Equation'',
         {\tt [nucl-th/9706029]}.

\bibitem{BRAATEN97}E.~Braaten and A.~Nieto, Phys.\ Rev. B {\bf 55}
    (1997) 8090;
      {\bf 56}    (1997) 14745.

\bibitem{HAMMER00}
H.-W.\ Hammer and R.J.\ Furnstahl,
Nucl.\ Phys.\ A{\bf 678} (2000) 277, {\tt [nucl-th/0004043]}.


\bibitem{NEGELE88}
J.~W.\ Negele and H.~Orland, {\it Quantum Many-Particle Systems\/}
    (Addison-Wesley, New York, 1988).

\bibitem{FETTER71}A. L. Fetter and J. D. Walecka, {\it Quantum Theory of
         Many-Particle Systems\/} (McGraw--Hill, New York, 1971).  

\bibitem{KOHN60}W.~Kohn and J.~M.\ Luttinger, Phys.\ Rev.\ {\bf 118}
  (1960) 41.       
 
\bibitem{HK64}P. Hohenberg and W. Kohn, Phys.\ Rev. {\bf 136} (1964)
B864.

\bibitem{FURNSTAHL01}R.J.\ Furnstahl, H.-W.\ Hammer, and N.\ Tirfessa,
 Nucl.\ Phys.\ A {\bf 689} (2001) 846, and references therein.

\bibitem{BRUUN98}G.~M.\ Bruun and K.~Burnett, Phys.\ Rev.\ A {\bf 58}
   (1998) 2427.

\bibitem{HU96}Y.~Hu, Phys.\ Rev.\ D {\bf 54} (1996) 1614.

\bibitem{OKUMURA96}K.~Okumura, Int.\ J.\ Mod.\ Phys.\ A {\bf 11} (1996)
    65.      
 
\bibitem{YOKOJIMA95}S.~Yokojima, Phys.\ Rev.\ D {\bf 51} (1995) 2996. 

\bibitem{COLLINS86}J.~C.\ Collins, {\it Renormalization}
  (Cambridge Univ.\ Press, 1986).

\bibitem{BLATT67}J.~M.\ Blatt,  J. Comp.\ Phys.\ {\bf 1}
  (1967) 382.
  
\bibitem{Furnstahl:2002gt}
R.~J.~Furnstahl and H.~W.~Hammer,
arXiv:nucl-th/0208058.
   
\bibitem{TELLER62}E. Teller, Rev.\ Mod.\ Phys.\ {\bf 34} (1962) 627.

\bibitem{FURNSTAHL99b}R.~J.\ Furnstahl and B.~D.\ Serot,
     Nucl.\ Phys.\ {\bf A671} (2000) 447.

\bibitem{HOLLAND01}M. Holland, S.~J.~J.~M.~F.\ Kokkelmans, M.~L.\
Chiofalo, and R.~Walser, Phys.\ Rev.\ Lett.\ {\bf 87} (2001) 120406.

\bibitem{TIMMERMANS01}E.~Timmermans, K.~Furuya, P.~W.\ Milonni, and
A.~K.\ Kerman, Phys.\ Lett.\ A {\bf 285} (2001) 228.

\bibitem{INAGAKI92}T.~Inagaki and R.~Fukuda, Phys.\ Rev.\ B {\bf 46}
  (1992) 10931.

\end{thebibliography}
\end{document}